# Next Generation High Brightness Electron Beams from Ultra-High Field Cryogenic Radiofrequency Photocathode Sources

J.B. Rosenzweig[1], A. Cahill[1], V. Dolgashev[2], C. Emma[1], A. Fukusawa[1], R. Li[2], C. Limborg[2], J. Maxson[1], P. Musumeci[1], A. Nause[1], R. Pakter[1], R. Pompili[3], R. Roussel[1], B. Spataro[3], and S. Tantawi[3]

[1]*Department of Physics and Astronomy, University of California, Los Angeles*
[2]*SLAC National Accelerator Laboratory, Stanford University*
[3]*Istituto Nazionale di Fisica Nucleare, Laboratori Nazionali di Frascati, Frascati, Italy*

**Abstract**
Recent studies of the performance of radio-frequency (RF) copper cavities operated at cryogenic temperatures have shown a dramatic increase in the maximum achievable surface electric field. We propose to exploit this development to enable a new generation of photoinjectors operated at cryogenic temperatures that may attain, through enhancement of the launch field at the photocathode, a significant increase in five-dimensional electron beam brightness. We present detailed studies of the beam dynamics associated with such a system, by examining an S-band photoinjector operated at 250 MV/m peak electric field that reaches normalized emittances in the 40 nm-rad range at charges (100-200 pC) suitable for use in a hard X-ray free-electron laser (XFEL) scenario based on the LCLS. In this case, we show by start-to-end simulations that the properties of this source may give rise to high efficiency operation of an XFEL, and permit extension of the photon energy reach by an order of magnitude, to over 80 keV. The brightness needed for such XFELs is achieved through low source emittances in tandem with high current after compression. In the XFEL examples analyzed, the emittances during final compression are preserved using micro-bunching techniques. Extreme low emittance scenarios obtained at pC charge, appropriate for significantly extending temporal resolution limits of ultrafast electron diffraction and microscopy experiments, are also reviewed. While the increase in brightness in a cryogenic photoinjector is mainly due to the augmentation of the emission current density via field enhancement, further possible increases in performance arising from lowering the intrinsic cathode emittance in cryogenic operation are also analyzed. Issues in experimental implementation, including cavity optimization for lowering cryogenic thermal dissipation, external coupling, and cryo-cooler system are discussed. We identify future directions in ultra-high field cryogenic photoinjectors, including scaling to higher frequency, use of novel RF structures, and enabling of an extremely compact hard X-ray FEL.

## I. Introduction

The introduction of fundamentally higher brightness electron sources, facilitated by the invention of the high-field radio-frequency (RF) photoinjector over 25 years ago [1][2], has changed the face of beam-based science. These sources have enabled the production of intense, cold, relativistic electron beams with ultra-fast time structures, that in the earliest days reached the picosecond scale, but now operate at the femtosecond level. Such high brightness electron RF photoinjector sources, based on short-pulse laser excitation of a photocathode embedded in a high-field RF accelerating cavity, have proven to be essential instruments in beam physics, enabling a range of high impact applications. These include the driving of very high gradient wakefield accelerators, in which a high-intensity, short pulse of electrons is used to excite high frequency, GV/m electric fields in plasma [3] or THz structures [4]. They also provide high brightness injectors for a variety of new ultra-fast, narrow spectrum light sources. This burgeoning class of instruments prominently includes the X-ray free-electron laser (XFEL) [5], which has revolutionized X-ray-based imaging since its inception, through the introduction of coherence in photon wavelength regions down to the Å level. With femtosecond pulses, the XFEL yields detailed information about the behavior and structure of atomic-molecular systems at their characteristic spatial and temporal scales, permitting so-termed ultra-fast, four-dimensional imaging. In all of these cases, the improved performance metrics of emittance and brightness are traceable to the order of magnitude increase in the electric field at emission compared to previous techniques.

The wakefield accelerator, which may enable GeV/m acceleration in future high energy electron-positron colliders, and the XFEL are large-scale instruments utilized in national laboratories. High brightness electron beams are also essential components in smaller, university lab-scale light sources, in the form of psec-resolution, quasi-monochromatic X-ray inverse Compton scattering (ICS) sources [6]. These sources do not produce coherent radiation, but permit very high-energy photon production, from the keV to MeV level, with relatively modest beam energy. Finally, high brightness electron beams having a few MeV kinetic energy find direct use in state-of-the-art imaging systems employing the electrons themselves. Indeed, ultra-fast relativistic electron diffraction (UED) and microscopy (UEM) [7][8][9] are emerging applications drawing significant attention from the structural dynamics imaging community.

In this list of high brightness electron beam applications, one stands out in terms of current and future impact — the central role played in creating the lasing medium for the XFEL, as typified by the LCLS [10]. The LCLS serves as a flagship and prototype of the 4[th]-generation of X-ray light sources [11][12][13], introducing ultrafast high-flux, coherent hard X-ray pulses. The enabling of self-amplified spontaneous emission (SASE) FEL [14] operation using an exponential gain regime based on high brightness electron beams has produced X-ray light sources having over ten orders of magnitude increase in peak photon spectral brilliance compared to preceding sources. These extremely bright, coherent light sources have introduced powerful, innovative methods in X-ray-based science [15]. The LCLS is proceeding to a significant upgrade at present[16], mirroring the current worldwide investment in the XFEL sector.

The connections between XFELs and the concepts of electron beam brightness and emittance are fundamental. One quantifies the five-dimensional beam brightness as $B_e = 2I/\varepsilon_n^2$, where $I$ is the peak beam current, and $\varepsilon_n$ is the normalized transverse emittance. Low emittance not only increases the brightness, but sets the minimum FEL wavelength achievable with a given be $\lambda_r > 4\pi\varepsilon_n/\gamma$, a condition now commonly termed the Pellegrini criterion [17,18]. In high gain FEL theory the unitless gain parameter $\rho$ is found to depend on the electron beam brightness as $\rho \propto B_e^{1/3}$. This parameter controls the exponential gain length, as $L_g \propto \rho^{-1}$ and, in the case of the SASE FEL, the efficiency, $\eta = \frac{U_{FEL}}{U_{e-}} \simeq \rho$. Here $U_{FEL}$ and $U_{e-}$ are the total beam energies in the photon and electron beams, respectively. Improvement in brightness is critical to the success of the X-ray FEL; without the order of magnitude increase in brightness achieved through a combination of improvements in RF gun realization and downstream beam dynamics methods which mitigating coherent instabilities, the LCLS would not have reached saturation in its 120-meter undulator [10]. Indeed, the need to traverse ~20 power gain lengths to attain saturation remains a major challenge for advanced schemes such as self-seeding.

In similar way, high beam quality is needed in future electron-beam-based imaging systems. The lowering of $\varepsilon_n$ plays an enabling role in the feasibility of UED, as the spread in electron beam angles must be smaller than the diffraction angle associated with the electrons' de Broglie wavelength. Peak current is also needed to permit a larger number of electrons within a certain measurement time (psec-to-fsec) to observe, *e.g.*, dynamic changes in material properties revealed through diffraction. The use of intense, ultra-relativistic electron beams in imaging microscopes is termed UEM. It has yet more demanding, beyond the state-of-the-art requirements on emittance and intensity performance, and its realization is attracting significant recent attention [19].

With the central role played by brightness in XFEL performance and other frontier applications, the search for methods that increase the brightness $B_e$ has taken on increased urgency. Given the excellent performance obtained in compensating the space-charge induced emittance components through judicious control of the beam's transverse plasma oscillations, efforts have recently turned towards reductions of the intrinsic cathode emittance (often termed the *thermal* emittance), through choice of materials and laser wavelength. The intrinsic five-dimensional beam brightness *at emission* is inversely proportional the beam's effective initial temperature $T_c$ [20],

$$B_e = \frac{2I}{\varepsilon_n^2} = \frac{2\pi J_{\max} m_e c^2}{k_B T_c}. \tag{1}$$

Here the parameters $k_B$ and $m_e c^2$ indicate the Boltzmann constant and the electron rest energy, respectively. In this definition the current $I$ is divided by the 4-dimensional transverse phase space volume $\varepsilon_n^2$, a ratio we have recast in terms of maximum current density at emission $J_{\max}$ (for a Gaussian beam distribution) and the effective electron emission temperature $T_c$. Both of these parameters can be examined for their possible improvement. For sub-psec emission from metallic surfaces $k_B T_c$, often termed the mean transverse energy (MTE), is near to the difference between the laser photon energy and the metal's $\phi_{\text{eff}}$. We note that this assertion concerns scenarios where the photocathode ambient *material temperature* $T_m$ (specifically, that of the electrons in equilibrium internal to the material) is negligible, an issue that must be revisited below. The approach to improving brightness that depends on significantly lowering the intrinsic emittance implies, in prompt-emission metallic photocathodes, a concomitant lowering of quantum efficiency. As such attempts to lower $T_c$, while progressing, have thus far produced only moderate increases in the brightness obtained from sub-psec-response photocathodes [21].

Equation 1 indicates a powerful and direct approach to increasing the electron beam brightness, through augmenting $J_{\max}$. As we shall see, this approach promises over an order of magnitude increase in $B_e$, obtained through significantly increasing the peak accelerating field at emission. This is enabled by cryogenic operation of the RF structure containing the photocathode. The potential advantages of high field operation are explicitly demonstrated from the expression for the maximum current density obtained from a photocathode in 1D space-charge limited flow (very short initial beam, or blowout, regime per the discussions in Ref. [22] and in Appendix A),

$$J_{z,b} \cong \frac{ec\epsilon_0 (E_0 \sin\varphi_0)^2}{m_e c^2} = \frac{I_0 (\gamma')^2}{4\pi}. \tag{2}$$

Here $I_0 = \frac{ec}{r_e} \cong 17 \text{ kA}$, $\varphi_0$ is the RF cavity phase when the laser impinges on the photocathode and $E_0 \sin\varphi_0 \equiv \gamma' m_e c^2 / e$ is the extraction field at this phase. The parameter $\gamma'$ is the initial accelerating gradient experienced by the electron normalized to its rest energy; $\gamma'^{-1}$ thus measures the distance needed to increment the energy by $m_e c^2$. As this expression gives the current limit in the 1D-limit longitudinal "blowout" regime of operation, we use the subscript *b* to indicate it. We can employ Equation 2 to estimate the associated intrinsic limit on peak beam brightness for this case of uniform emission, similar to the discussion of average beam brightness in Ref. [23],

$$B_e \cong \frac{ec\pi\epsilon_0(E_0\sin\varphi_0)^2}{k_B T_c} = \frac{I_0(\gamma')^2}{4(k_B T_c/m_e c^2)}. \tag{3}$$

This scenario also gives a potential path to emittance growth minimization, in the sense of enabling robust phase space dynamics due to the formation of a nearly uniform ellipsoid of charge, to obtain self-fields linear in offset in all three spatial directions. As we shall see below, however, despite this advantage, the blowout approach is limited in its effectiveness by the associated introduction of a correlated energy spread that can interfere with the emittance compensation process [24].

If the laser is not transversely shaped to specifically produce the desired ellipsoid, but instead uses a transversely flat laser intensity, the beam expands to approach a uniform cylindrical distribution with inherent brightness remaining as in Eq. 3. This distribution is historically important, as it was the favored form in the original emittance compensated designs for *e.g.* the LCLS [29]. In this regard, it is also useful to explicitly write the total limiting current in the quasi-1D, longitudinal motion-dominated blowout regime. Assuming transversely uniform emission and the formation of a cylindrical beam, for the purpose of a later comparison with the 2D, long beam case, we have

$$I_b \cong J_{z,b}\pi R^2 = \frac{I_0}{4}(\gamma' R)^2, \tag{4}$$

where $R$ is the radius of the beam edge.

It is notable that by significantly increasing the launch field $E_0 \sin\varphi_0$ one accesses the brightness advantages indicated by Eq. 3. One may attempt to increase this field through various approaches, including operation with very short impulses of RF power to avoid breakdown due the effects of pulsed surface heating and large electric fields. To this end, it is attractive to use higher RF frequency $f_{RF}$, as the time needed for inserting and extracting power into and out of standing wave RF devices scales as $\tau_F \propto f_{RF}^{-3/2}$, where $\tau_F$ traditionally indicates the fill time. Indeed, while peak electric fields in an S-band photoinjector may reach ~160 MV/m, similar X-band structures have been operated at yet higher fields [25].

The advantages of higher frequency operation are challenging to realize, however, as the shortness of the wavelength for cases below S-band implies that $\varphi_0$ may be notably less than $\pi/2$. This problem is quantified through the parameter $\alpha_{RF}$, the normalized vector potential amplitude associated with the RF field [26], where the free-space RF wavelength $\lambda_{RF} = c/f_{RF}$. Using $\alpha_{RF}$ one can estimate the dependence of $\varphi_0$ on $E_0$ o. The optimal phase slip $\Delta\varphi_0 = \varphi_\infty - \varphi_0 = \frac{\pi}{2} - \varphi_0$ for the initial cell of length $\lambda_{RF}/4$ is approximately given [26] by the transcendental relation $\left(\frac{\pi}{2} - \varphi_0\right)\sin\varphi_0 \simeq \frac{1}{2\alpha_{RF}}$. For S-band with $E_0 = 120$ MV/m and an initial cell of length $\lambda_{RF}/4$, as in the first generation high gradient RF guns [27], one may inject at $\varphi_0=75°$.

We note, however, that in more recent photoinjector designs [28] an initial cell 1.2 times longer than the original $\lambda_{RF}/4$ is utilized to aid in optimizing the transverse dynamics [29] through increased RF focusing effects. As standing-wave photoinjectors commonly employ π-mode ($\lambda_{RF}/2$ cell length) structures, this scenario is referred to as 0.6-cell, meaning $0.6(\lambda_{RF}/2)$. The additional cell length implies a smaller initial launch phase. Further, in order to counter the pulse lengthening due to space-charge effects in the 0.6 initial cell case by providing longitudinal focusing, LCLS operation requires a launch phase at 120 MV/m of $\varphi_0=30°$ in practice. In this scenario the beam experiences only 60 MV/m at emission. On the other hand, by *shortening* the initial cell from $\lambda_{RF}/4$ , one may effectively launch near RF crest [19]. This can be enhanced by reducing the initial cell length, but at a cost in transverse beam quality, particular for higher $Q_b$ cases.

In studies of photoinjector operation in X-band, at $f_{RF}$ =11.424 GHz ($4f_{RF}$ of the LCLS S-band case), fields of >200 MV/m have been demonstrated. However, it is difficult to take advantage of these fields as $\alpha_{RF}$ is small, near 0.8 in this case, due to the significant shortening of the RF period. As such, one may not easily simultaneously improve on the launch field and transverse dynamics obtained at, *e.g.,* the LCLS using present techniques. Also, in X-band the emittance compensation focusing solenoids are very challenging [30]. Indeed, the brightness obtained in X-band photoinjectors has not yet yielded significant improvement despite use of higher field amplitude [30]. We note, however, that operational challenges in X-band are mitigated in C-band, and brighter beams may be obtained, as discussed below. Because of the large value of $\alpha_{RF}$ used, however, the optimized emittance predicted in the S-band case we examine in detail is lower for the same electric field amplitude $E_0$.

We can now introduce the experimental motivation of the initiative described in this paper. While some increases in brightness may be achieved by optimizing current methods, to reach significant increases a change in approach is needed. We propose here a new paradigm for photoinjector realization, profiting from successful work in the development of cryogenically-cooled Cu RF structures recently undertaken by SLAC and UCLA. Remarkably, in tests on X-band structures operated at 45 °K, an enhanced quality factor $Q_0$ and significantly higher fields, corresponding to nearly $E_0$ =500 MV/m surface fields before breakdown, have been demonstrated [31]. These advantages arise from the diminished surface dissipation associated with the anomalous skin effect (ASE), the improved material strength in Cu at cryogenic temperatures, as well as a diminished coefficient of thermal expansion, all of which aid in preventing mechanical stress due to pulse heating. We note that while exploiting the lower dissipation [32] to permit *high repetition rate,* high duty factor injectors at increased gradients has been examined in previous investigations [33], the use of cryogenic RF structures to achieve very high fields in lower duty factor guns, with their attendant benefits in beam brightness, is the key initiative newly introduced in this paper. Thus, based on the recent progress in pushing the frontier of attainable fields in RF structures, we analyze in detail a scenario that seeks to profit from cryogenic operation of copper cavities, as applied to an advanced RF photocathode gun. This discussion concentrates on the possibilities of developing an S-band, 1.45 cell photoinjector gun operated near 27 °K, with $Q_0$ enhanced by a factor of up to five and, most critically, a peak electric field on the cathode of at least 250 MV/m. After the discussion of this optimized S-band system, we return briefly to discuss some potential practical advantages found in extending the concept of an ultra-high field photoinjector to higher frequency.

In the S-band scenario, one may reach an unprecedented level value of $\alpha_{RF}$ ~4. In this case, the value of $\sin\phi_0$ approaches unity [19,26] for the assumed 1.45 cell π-mode structure. Thus, the launch field is four times larger than that currently used in the LCLS [23], and the brightness is predicted to be increased 16-fold for 1D space-charge limited flow limit; this number is modified somewhat by 3D effects. Further, as the accelerating field is twice as large as presently used, the beam exits the photoinjector with approximately double the energy of present devices. This yields advantages in handling related deleterious collective effects in beam transport.

We note that additional enhancement of brightness may be expected through lowered intrinsic emittance, that is, through a decrease in emitted effective electron temperature. This issue is also affected by the photocathode material temperature, as well as improvement in the vacuum environment at low temperature, the laser photon energy, and the Schottky effect, etc., as discussed below. With smaller emission areas and reduced $T_c$ compared to current parameters [28], the order-of-magnitude improvement in brightness $B_e$ should strongly benefit future X-ray FELs, with much smaller gain length $L_g$, and a concomitant increase in the power efficiency. These improvements also positively impact self-seeding schemes [34][35], which are based on manipulations of the electron and radiation beams over many gain lengths.

To profit from much higher brightness in the X-ray FEL context, one still must optimize the initial space-charge dominated beam dynamics through emittance compensation. Further, once this emittance-minimized beam is in hand, it must then be compressed from sub-100 A peak current to many kA. With initial emittance smaller by an order of magnitude, this implies confronting new challenges in the control of collective effects, particularly longitudinal space charge and coherent synchrotron radiation (CSR), during transport and compression. The issue of CSR has been addressed previously in computation [36,37] and experiment [38], in the context of proposals to use very low charge and emittance beams as a path to achieving single spike, sub-fsec SASE pulses. Single spike operation is attractive, as it may extend nonlinear optical techniques used in ultra-fast chemistry and atomic-molecular physics to the X-ray region [39]. Experimental work in this context was performed at 20 pC, and achieved 2 fsec rms pulse length, or ~8 kA peak current, but at the cost of growth from 0.14 to 0.4 mm-mrad in $\varepsilon_n$ in the final compression chicane's bend plane. This growth is a limitation that must be considered and mitigated when considering much brighter, low emittance beams.

The scenario explored in Refs. 36-38 concerned use of small charge pulses, but with standard methods of pulse compression. Given the presently understood limitations of these methods, here we examine a promising alternative, showing the results of start-to-end simulations of an XFEL employing a novel approach to final bunch compression and lasing. This technique is termed enhanced self-amplified spontaneous emission (ESASE) [40]. We demonstrate that with strongly lowered emittance, the performance of an FEL using ESASE is greatly improved, in the sense of much shorter gain length and higher efficiency. Further, there are new capabilities accessed with such small emittance. In particular, FEL wavelengths an order-of-magnitude smaller than present LCLS operations are permitted. We illustrate this with simulation of a compact, 80 keV photon-energy X-ray FEL employing an advanced short period undulator [41] and a beam at LCLS energy. In this case saturation occurs within 20 m, due to very high beam brightness.

Finally, we note that there is a strong demand in the FEL science community for higher photon flux per pulse, to reach the level needed for imaging large systems such as protein molecules [42]. These applications require peak powers in the multi-TW range, contained in 25-100 fs pulses, to permit imaging before the destruction of the target. This approach utilizes high currents, at the 4-5 kA level, to permit self-seeding, saturation, and tapering within a reasonable length, *i.e.* less than the 140 m foreseen for LCLS II undulators. Tapering is essential to this approach, as one must extract over 5% of the beam power as X-rays. We also note that the efficacy of tapering is directly enhanced by having a higher power X-ray pulse, due to enhanced brightness, at the onset of saturation – more radiation field is available to decelerate the electron beam trapped in the FEL's ponderomotive potential [43][44]. In this paper, we show that for LCLS-like cases that notably higher saturation powers when the very low emittance photoinjector in tandem with ESASE bunching are employed. With a more sophisticated understanding of the tapering process, recent proposals have demonstrated that a beam charge of 100 pC, as we examine in this paper, may be sufficient to achieve the photon flux demanded by single molecule imaging.

The structure of this article is as follows. We first review recent advances in peak field achieved in RF structures operated at cryogenic temperatures in both X-band and S-band. Based on this discussion, we take an expected peak field performance of $E_0$=250 MV/m on the photocathode in a cryogenic S-band 1.45 cell RF gun. This RF scenario is chosen due to its relatively straight-forward implementation in existing photoinjector systems, and to maximize the injection field and the associated beam brightness. We give a detailed discussion of the beam dynamics in this and related scenarios. A schematic layout of this cryogenic gun system is displayed in Fig. 1; not shown are post-acceleration sections that are found in XFEL injectors.

We concentrate first on a high beam charge ($Q_b$) cases aimed at XFEL application [10] in which operation in the blowout regime is assumed. This serves to illustrate the enhancement of the current $I$ in the 1D limit, and it also shows the problems that induced energy spread gives in achieving emittance compensation. To understand how to mitigate this problem, we examine a low charge, low emittance case directed towards UED and UEM application [19], where "cigar-beam" emission is employed, and a factor of 50 improvement in brightness over existing injectors is found. We discuss the impact of this level of brightness on UEM temporal-spatial resolution.

Using the results obtained in the low charge study, we return to the optimization of the dynamics for higher charge beams in cigar-beam-like cases. This is done by examining a C-band example where the beam, RF cavity, and focusing parameters are scaled with RF wavelength [45] from a re-optimization of the LCLS photoinjector [46]. Using the C-band operating point to give direction to S-band ultra-high field operation – in particular in understanding necessary modifications to the placement of the post-accelerating linac – we examine cases where FEL-quality electron beams are produced in simulation with $\varepsilon_n$ = 0.036 mm-mrad at 100 pC, representing over an order-of-magnitude increase in both $\varepsilon_n$ and $B_e$ over the state-of-the-art. This beam is utilized in start-to-end simulations [47] using the LCLS beamlines and undulator, along with ESASE. Significantly enhanced FEL power and efficiency are found. We then examine a case which uses a short-period undulator to produce 80 keV X-rays, as are needed for the MaRIE project.

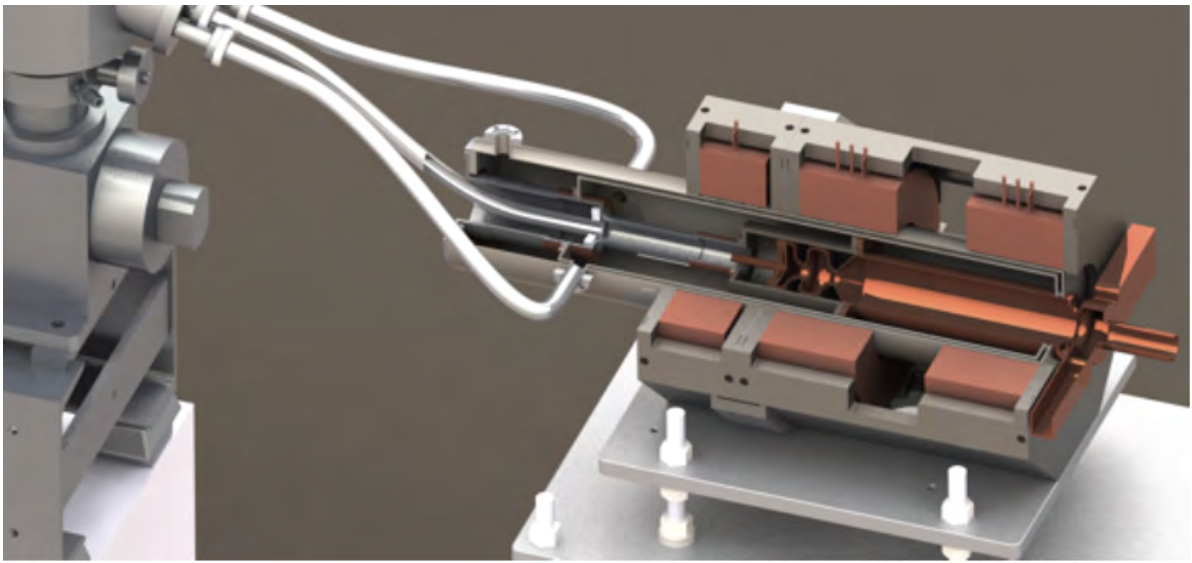

**Figure 1. Cryogenic, very high field S-band photoinjector, with 1.45 cell Cu gun structure (center) externally coupled to waveguide through a mode-launcher scheme (far right). Also shown: cryostat envelope and liquid neon-based cryo-cooler (far left), mounting equipment, emittance compensation solenoid (surrounding RF structure).**

Beyond the improvements expected from the use of very high fields, we discuss the diminishing of the intrinsic emittance expected at low temperature. We further review relevant aspects of the proposed RF design, including cavity shape and length, as well as optimization of the external coupling system. In this context we review the demands placed on the cooling system, and discuss their solution, illustrating the conceptual layout of the integrated system. We then revisit the possibilities for extending this approach to shorter RF wavelengths.

## II. Cryogenic operation of radio-frequency structures

Recent research aimed at improving the accelerating gradient in normal conducting RF structures, has made remarkable progress. A key finding is that cryogenic operation of Cu structures permits much higher breakdown thresholds. In experiments with short standing wave (SW) structures it is found that, after initial conditioning, and the breakdown rate is reproducible for structures of the same geometry and material. Further, the breakdown rate depends critically on the peak magnetic fields which give rise to pulsed heating and related phenomena [48], in combination with the peak surface electric fields [49]. In this regard, recent studies show that the breakdown rate correlates with the peak surface vale of a modified Poynting vector [50]. A current hypothesis seeks to explain the statistical behavior of RF breakdown in accelerating structures through generation and movement of dislocations under stresses created by RF electric and magnetic fields [51]. Resistance to this movement is predicted to improve by use of material with greater yield strength, *e.g.* Cu alloys. Further, the yield strength is systematically enhanced at cryogenic temperature even in high purity Cu, and the coefficient of thermal expansion is lowered, giving much smaller mechanical stress. Indeed, recent studies carried out at SLAC on both harder Cu alloys (CuAg) and cryogenic Cu have given the desired results — dramatically higher surface electric fields are achieved before breakdown, as is summarized in Figure 2.

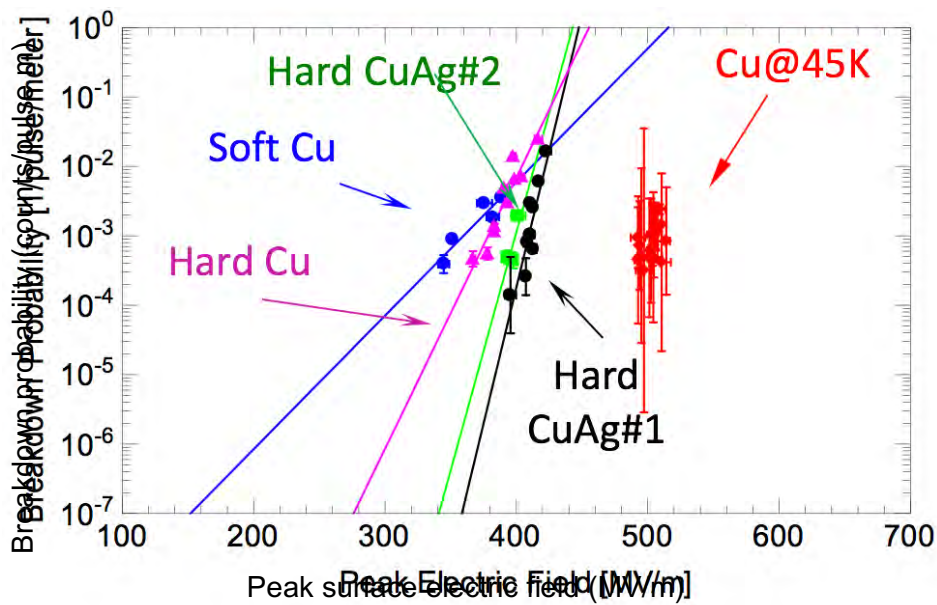

**Figure 2. Breakdown probability in per pulse-meter of structure length as a function of peak surface electric field in single cell X-band accelerating structure tests. The introduction of a harder alloy (CuAg, two different samples, indicated as #1 and #2) improves the breakdown as predicted; the effect of operation at 45 deg K is more dramatic, permitting surface electric fields to a threshold at 500 MV/m. From Ref. 31.**

Figure 2 illustrates the results that have been obtained from experiments performed on single cell

X-band accelerating structures [31] of a modified pillbox design similar to cavity shapes used in photoinjectors. The dependence on peak surface electric field of the observed breakdown probability per pulse (per meter of structure length) shows qualitatively and quantitatively different behavior. The introduction of a harder alloy improves the breakdown as predicted, while the effect of operation at 45 °K is more dramatic still, permitting surface electric fields up to a sharp breakdown threshold near 500 MV/m. In the case of cryogenic operation, there is effectively no breakdown below this threshold, in contrast to the room temperature cases. This advantageous change in performance is due to the combined effects of increased yield strength, and to the above-noted lowering of surface heating due to the diminished surface resistivity.

High brightness photoinjectors, as noted above, have been generally operated in S-band, as this optimizes considerations of peak field, stored energy, wakefields, physical aperture, and wavelength dependence of RF focusing and longitudinal emittance minimization. In addition, any improvements to existing injector systems would be much more technologically feasible when utilizing the same RF power and timing system. As such, our detailed example of the first use of cryogenic copper in photoinjectors is presented below in S-band. We later examine some possible advantages of using of cryogenic copper in a high-field, C-band RF photoinjector.

### III. Beam dynamics: operation in the blowout regime

The scaling of the current density at emission in the 1D limit discussed above indicates strong improvements are possible at high fields, with brightness having a quadratic dependence on launch field in the blowout regime. As seen below, the peak $J_{\max}$ and $B_e$ at emission scales as such for high-$Q_b$ cases, as needed for both very high power FEL and wakefield acceleration applications. In order to explore the possibilities associated with the strong scaling of $B_e$ with launch field, we first address beam dynamics issues arising in the 1D blowout regime limit. We take as an illustrative example a case with significant charge $Q_b\sim$125 pC.

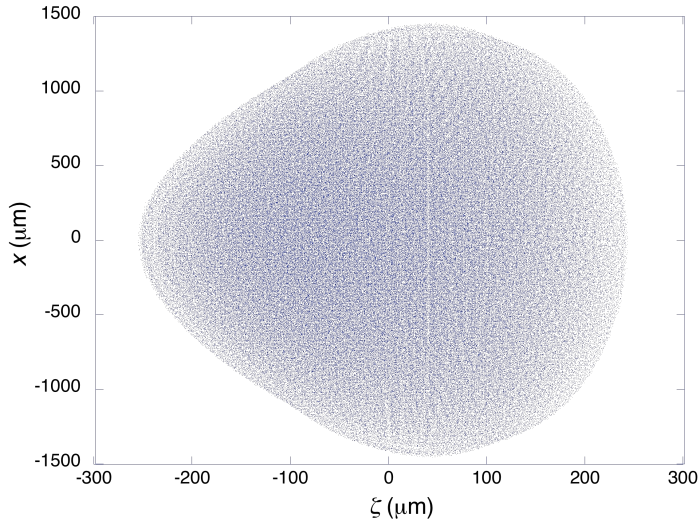

**Figure 3.** Approximation of ellipsoidal distribution formed after 250 MV/m peak field RF photoinjector (downstream of photocathode z=1.5 m) in 125 pC blowout regime case.

The beam dynamics analysis leading to the estimate of current and brightness limits given in Eqs. 2 and 3 is found in the Appendix. That discussion self-contained, but it is worth noting that in addition to a current limit, there is a maximum $Q_b$ that can be extracted from a photocathode in the 1D limit [23,52], which is $Q_{b,\max} = \varepsilon_0 E_0 \sin\phi_0 \pi R^2$ for a radially uniform distribution. This implies, in terms of the current limit of Eq. 4, that $Q_{b,\max} = I/(\gamma' c)$, or that the limiting pulse length is $T_{\max} = (\gamma' c)^{-1}$. In practice, as discussed in the Appendix, degradation of the current distribution is seen well below $Q_{b,\max}$, and thus one should operate at $Q_b < 0.2 Q_{b,\max}$, to obtain approximately uniform current $I_b$.

For ~250 MV/m launch field, the maximum current using, for example, a 315 μm hard-edge-radius distribution the limit on $I_b$ is ~100 A. This 1D prediction is of course approximate, due to the 3D nature of the pulse expansion. We concentrate first on the example of $I_b$ =100 A, to draw connection to the original LCLS photoinjector design, which remains point of reference in the electron source field [53]. Specifically, the original proposal for the LCLS photoinjector employed approximately constant laser intensity inside of a cylindrical temporal-radial boundary, launching a nearly uniform cylinder of charge from the photocathode [28]. To compare this to an equivalent blowout regime case, we choose a transverse laser distribution corresponding to the "half circle" distribution, with intensity $\sim[1-(r/R)^2]^{1/2}$, as well as a laser pulse much shorter than the eventual length beam after longitudinal expansion yields a nearly uniformly-filled ellipsoid of charge [22][54][55]. This scheme produces a maximum current in the longitudinal space charge-dominated limit, and linear fields leading in principle to good emittance preservation – albeit only up to a certain $Q_b$ – and excellent compressibility [56][57][58]. The deviations from ellipsoidal shape displayed in Fig. 3 are due to problems arising when one approaches $Q_{b,\max}$.

Table 1. Parameters for blowout regime beam dynamics simulation.

| Laser pulse length | 35 fs FWHM, |
|---|---|
| Laser spot size (cut transverse Gaussian) | Hard edge at 262 μm, 1.6σ, (120 μm rms) |
| RF gun format | 1.45 cell π-mode standing wave |
| Peak cathode electric field | 250 MV/m |
| Launch phase | 82 degrees |
| Focusing solenoid (SPARC-type) field | 5.4 kG |
| Post-acceleration linac average field | 20 MV/m |

For an example of the beam evolution in this system we examined through GPT particle simulations [59] a case with the parameters summarized in Table 1. The photoinjector is followed by a 30 cm long solenoid with ~0.5 T peak field, and employing a design in use in numerous injectors worldwide. This magnet focuses the beam into a post-accelerating linear accelerator (linac) section 3-m in length that begins at $z$=1.5 m downstream of the photocathode, as found *e.g.* in the LCLS. This linac also has solenoid focusing superimposed, as has been introduced at the SPARC Laboratory [24] at INFN-LNF. These external geometric attributes are thus representative of current techniques. To operate this scenario in the blowout regime, we use a 35 fs FWHM laser pulse, with a specially tailored transverse distribution, a Gaussian cut at $R$=1.6σ, in this case 262 μm. This form allows approximation of the half-circle distribution, with an initial rms size transverse beam size in Cartesian projection of 120 μm. The intrinsic emittance is included at the level of 0.54 (mm mrad)/mm rms (MTE $k_B T_c$=0.15 eV) at launch.

Figure 3 displays the beam spatial distribution after longitudinal expansion during the initial acceleration in the RF photoinjector, giving a near-uniformly filled, approximately ellipsoidal distribution. The longitudinal phase space bears evidence of this expansion, with a large positive, nearly linear chirp displayed in Figure 4. The linearity of the chirp indicates good prospects for longitudinal compressibility, as needed for, *e.g.*, wakefield applications. It can, however, provoke problems in the transverse dynamics, as discussed below. We note that with this RF geometry and the high field used, that the final median beam energy is over 10.5 MeV after the RF photoinjector, or nearly double that of current devices. After post-acceleration to 160 MeV the peak current at injector exit is 100 A, as in the original LCLS design [29].

The transverse beam envelope evolution along the beamline direction *z* is shown in Figure 4 which displays similar behavior to present emittance-compensated RF photoinjector systems. The associated emittance evolution is also shown in Figure 4. The minimum shown at the waist (*z*=10 m downstream of the photocathode) is near to 0.2 mm-mrad, with a 0.16 mm-mrad slice emittance. This example shows notable room for improvement in the compensation process, as the intrinsic emittance for the beam launched here (~0.1 mm-mrad) was not reached. Nevertheless, the beam brightness here is much higher than the LCLS design, which called for a $\varepsilon_n$= 1.1 mm-mrad. The design brightness in this example of an ultra-high field RF photoinjector, even in this imperfect case, is increased by a factor of over 30 over the original LCLS design. This comparison does not consider recent advances in injector design, however.

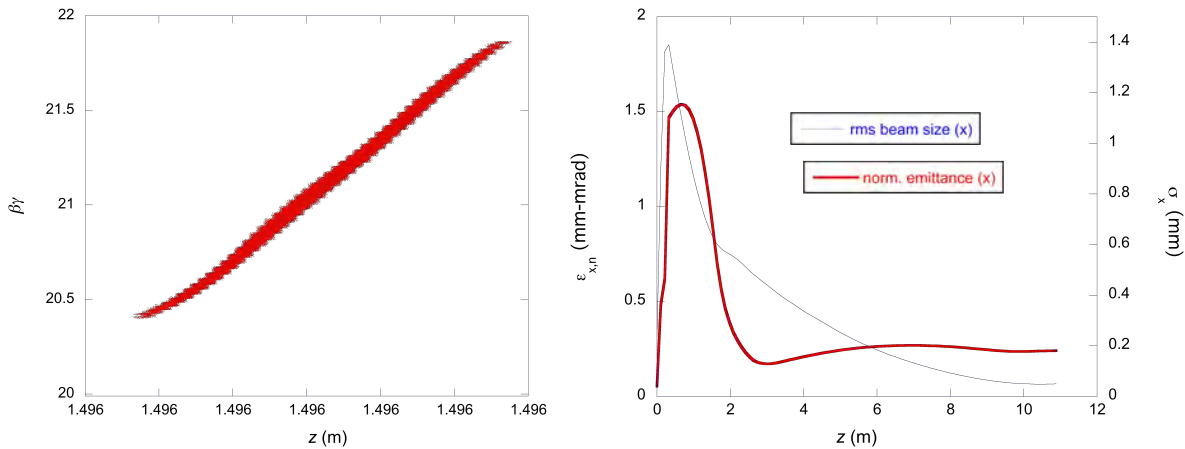

**Figure 4. (left) Longitudinal phase space for 125 pC beam distribution shown in Figure 3, after 250 MV/m RF photoinjector. (right) Associated transverse rms beam envelope and normalized emittance evolution.**

The beam dynamics are evidently not optimized in this blowout regime example. There are two reasons for this. First, this scenario for emittance compensation is entirely new, with a beam energy after the gun twice as large as the well understood, S-band, LCLS-like scenario. We have not changed the geometry of either the acceleration or the focusing schemes in this example, however, and these choices must be revisited. Second, the energy spread is large after the gun (*cf.* Ref. 28), where strong focusing is applied to obtain emittance compensation. This energy spread gives rise to chromatic aberrations that raise the final emittance after compensation. This blowout limit remains of interest for producing low emittance beams with high current directly out of the injector, as may be needed for wakefield acceleration. For FEL and UED/UEM, one must improve $\varepsilon_n$.

Thus, even though current density $J_{\max}$ and thus the brightness $B_e$ immediately after beam emission are enhanced in longitudinal blowout regime, other approaches to emittance minimization must be considered that do not necessarily operate near the one-dimensional space-charge limit. These options include the transverse blowout, or cigar-beam, regime where the space-charge induced motion after emission is primarily radial. We proceed to reviewing this regime.

## IV. The cigar-beam regime

The 1D limit discussed above illustrates the current scaling arising from longitudinal self-forces. The beam's longitudinal expansion in the blowout regime, resulting from use of an ultra-short laser pulse, can produce the desirable uniform ellipsoidal beam distribution, but can also yield a beam with excessive energy spread. As such, we examine the alternative use of cigar-shaped beams [60] in which the beam spatial dimensions at launch obey $L_z \gg R$, as opposed to the blowout regime, where $L_z \ll R$. In determining these conditions, the value of $L_z$ is evaluated, or for constant current emission lasting a time $\tau$, $L_z = (\gamma'/2)(c\tau)^2$. In the case $L_z \gg R$ the induced surface charges at the cathode spread out transversely, and the decelerating fields that cause pulse lengthening and ultimately virtual cathode formation (e.g. when $Q_b = Q_{b,\max}$ in the blowout) are diminished. In this regime one may launch a beam with increased charge emitted per unit area, and thus minimize the emittance at emission. Further, the beam in this regime does not dramatically lengthen, and one may obtain results, in terms of peak current and brightness, that are predicted for certain parameters to be quantitatively improved over the blowout regime.

To investigate this possibility, we refer to the Child-Langmuir-like analysis introduced in Ref. 60. The peak current obtained in the cigar beam limit (with subscript $c$ indicating cigar regime) is

$$I_c = I_0 \frac{\sqrt{2}}{9} \left[\frac{eE_0 \sin\phi_0 R}{m_e c^2}\right]^{3/2} = I_0 \frac{\sqrt{2}}{9} [\gamma' R]^{3/2}. \tag{5}$$

Using both Eqs. 4 and 5, and assuming the same radially uniform distribution to a hard-edge radius $R$, the ratio of cigar-to-blowout current is

$$\frac{I_c}{I_b} = \frac{4}{9} \left[\frac{2}{\gamma' R}\right]^{1/2} \simeq \frac{0.63}{(\gamma' R)^{1/2}}. \tag{6}$$

Thus, for a large enough beam size $R$ or field $E_0$, the advantage in initial current will be found in the blowout regime. Note that this is consistent with obeying the limiting $Q_{b,\max}$. In our example of the moderate beam charge case discussed above, however, assuming a rms equivalent cylindrical beam size (220 μm), the factor $(\gamma' R)^{-1/2}$=3 and $I_c \sim 2\, I_b$.

On the other hand, one must maintain a cigar aspect ratio while holding $R$ constant to access the cigar-beam regime. Exploring the S-band scenario further, we take the practical limit on the pulse length as $\tau$ =10 ps (flat-top profile), similar the LCLS design, as well as LCLS re-optimization case discussed below, to avoid degradation of the longitudinal phase space and concomitant chromatic aberrations due to the beam's lengthy extent in RF phase. In other words, the scaling law $I_c$ for may be applied only subject to geometric concerns. To quantify this issue, one may set $R = L_z$ to find the definitive violation of the cigar assumption, that is $R = (\gamma'/2)(c\tau)^2$, and then use the practical limit $c\tau = \lambda_{RF}/36$ (10° phase extent). Together with Equation 5, we arrive at the value of $Q_b$ that one must operate well below in order to launch a cigar-beam,

$$Q_{b-c,\max} = \frac{I_0}{18c}\gamma'^2 \left(\frac{\lambda_{RF}}{36}\right)^2. \tag{7}$$

For our S-band scenario with $\lambda_{RF}$ =0.105 m, and assumed 250 MV/m operation, this limiting cigar-beam charge is $Q_{b-c,\max}$ =50 pC. One should thus take the cigar-beam regime scaling to be approximately valid for high brightness beam production when $Q_b$<10 pC. In accordance with the scaling above, we indeed we will find that the cigar-beam regime offers advantages in producing the highest brightness beams at low charge $Q_b$. As will be clarified in subsequent discussions, longer beams with smaller radial extent tend to produce more optimized emittance compensation (avoiding space-charge induced energy spread), and we will also utilize such beams, that are not quite in the cigar-beam limit, in emittance-optimized moderate beam charge scenarios for FEL.

Beams with low current may be in principle be compressed, and so to compare performance one may introduce a 4D brightness, which we indicate as $B_{4D} = 2Q_b/\varepsilon_n^2$, the time integral of $B_e$. This quantity was indeed used in Ref. 23 when discussing *average* brightness, which is $B_{4D}$ multiplied by the pulse repetition rate. As one may compress the beam, assuming the compression process does not significanltly increase the emittance, the quantity $B_{4D}$ is also taken as an important figure of merit in evaluating beam quality.

The motivation for longer pulse length $\tau$ in the cigar-beam regime is made explicit from the scaling of $B_{4D}$ in the cigar-beam regime. This scaling may be explicitly written as

$$B_{4D} = \frac{2Q_b}{\varepsilon_n^2} = \frac{I_0 \tau \gamma'^{3/2}}{9\sqrt{2R}} \left(\frac{k_B T_c}{m_e c^2}\right), \qquad (8)$$

showing the merit of using a large $\tau$ and small $R$. We illustrate this regime through an example.

### V. Low charge, extreme low emittance beams in the cigar regime

To illustrate the relative advantages of the cigar-beam regime, we first concentrate the discussion on the case of low charge $Q_b$, as has been studied previously in the context of UED and UEM [19]. We assume the emission of a 2 ps full-width beam distribution having a hard radial edge at 20 $\mu$m, which reduces $Q_b$ to 1.6 pC. In this case, $L_z \gg R$ and the factor $(\gamma' R)^{-1/2}=10$; the estimated peak current in this scenario is six times than that possible in the blowout regime. As we launch a beam in this example with current slightly below $I_c$, no significant bunch lengthening is foreseen.

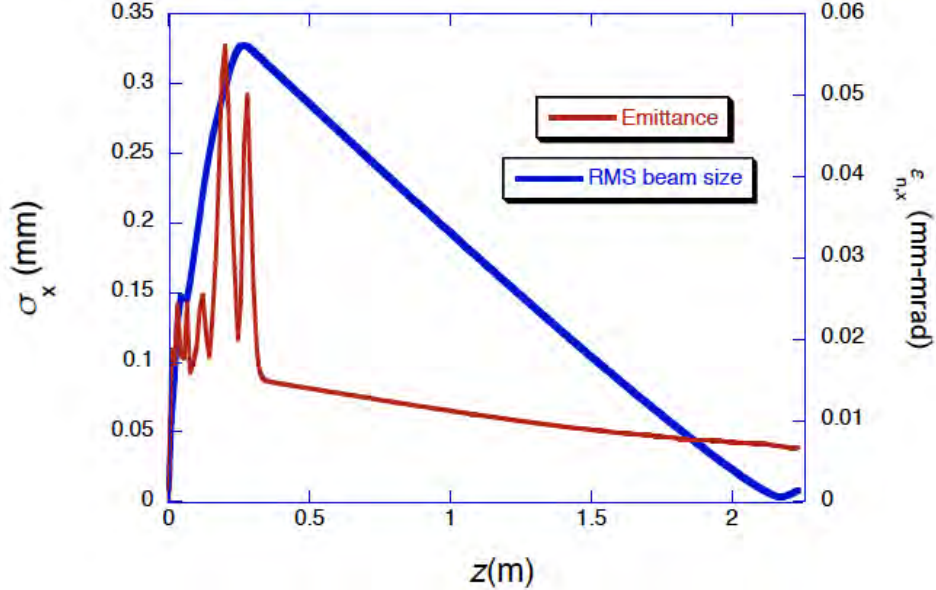

**Figure 5. Beam emittance evolution in low-charge, cigar beam case, showing an emittance compensated down to $=0.005$ mm-mrad level with 1.3 A peak current.**

The cigar-beam regime has an analogous process that found in the blowout regime, in which the laser temporal profile may be adjusted to give a half-circle intensity profile in *t*. The emitted beam then expands *radially* to give a nearly uniform-density ellipsoidal electron density distribution. The rapid radial expansion after photoemission plays a similar role to longitudinal expansion in the blowout regime, maximizing current at low charge while avoiding excessive energy spread.

Simulations using GPT have been performed to show the advantages of this mode of operation at low charge, using the parameters given above. The results of this numerical study, in which the beam focusing is optimized to produce a small spot $z$=2.2 m downstream of the photocathode, are shown in Figure 5 and Figure 6. The beam envelope arrives at an emittance-compensated waist where the original thermal emittance is recovered, near $\varepsilon_n$ =5 nm-mrad. Further, the peak current $I$=1.3 A. Comparing these results to a similar discussion in Ref. 19, we find that $\varepsilon_n$ has been halved, while the current has been enhanced by 13. In all, the beam brightness is increased by ~50.

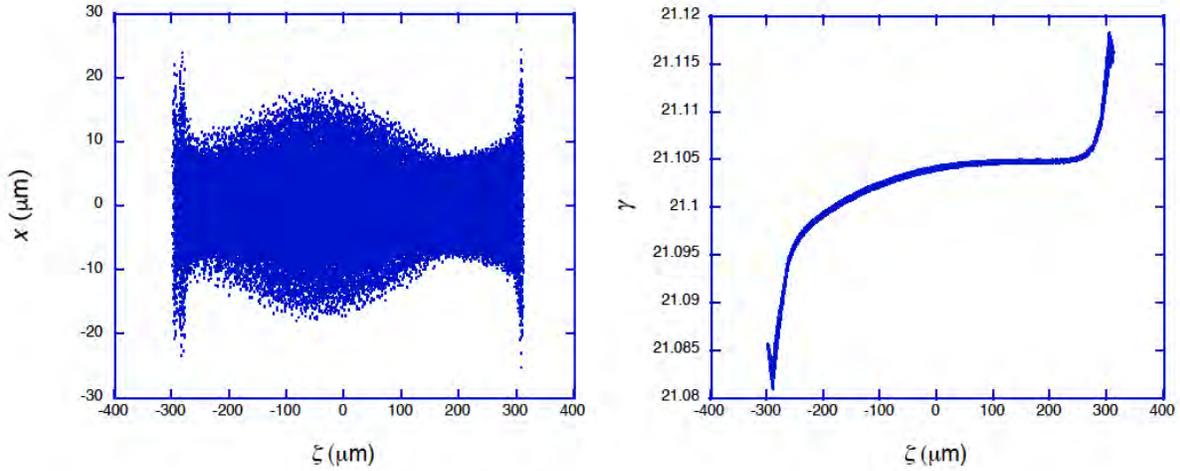

Figure 6 (left) Spatial beam distribution at emittance compensation minimum, in low-charge cigar beam case, at $z$=2.2 m, where the uniform distribution launched at photocathode is nearly recreated near the emittance compensation waist. (right) Longitudinal emittance for this distribution.

According to the discussion of the bright-field imaging process for microscopy included in Ref. 19, one should control the energy spread to the level of $10^{-4}$ to produce a measurement with 10 nm resolution. This is accomplished in the scheme of Ref. 19 by use of an X-band cavity to remove most of the correlated energy spread. It is notable that the same effect is obtained here by the exploitation of longitudinal space-charge forces, which produce a similar result. This is possible because of the increased current and tight electron beam focusing used. Further, the reduction of $\varepsilon_n$ and the increase in beam energy combine to permit better image contrast. As the beam charge is nearly the same in our case as in Ref. 19, the increase in *I* is due to a beam shortened by an order of magnitude; instead of ~10 ps temporal resolution, one reaches ~1 ps. Examining intensity dependent effects, the beam integrated flux is taken to be the same, and so sample damage is equivalent and has been evaluated as ignorable. Finally, the electron-electron interactions after the sample may cause degradation of the image formed. The macroscopic space charge forces scale as $I/\gamma^2$, and are thus slightly higher in this case; a re-optimization of the beam size may be necessary to obtain similar resolution. The same is true of microscopic space-charge (scattering) effects – they are nearly equivalent to the case discussed in Ref. 19, and may be further optimized by adjustment of the beam sizes and angles at the sample. In total, one may foresee development of an ultra-fast relativistic electron microscope with an order of magnitude faster time resolution than the previously proposed state-of-the-art.

With this promising scenario introduced for electron imaging applications, we next examine the use of dramatically higher brightness beams from high field photoinjectors in driving X-ray FELs. To this end, in the following sections we present studies of the optimization of higher charge – few 100's of pC – beams produced in the cryogenic photoinjector that enable qualitative leaps forward in XFEL performance.

## VI. Scaling the current state-of-the-art source to higher field, higher frequency

The optimization of emittance compensation remains, after more than 25 years of work, an active area of investigation. There have been significant clarifications obtained through computation and theory that illuminate the performance of the existing family of split photoinjectors, which use two-cell guns having $\alpha_{RF}$<2, with solenoid focusing that control the beam envelope – and thus transverse plasma oscillations) before injection into a post-acceleration linac. These studies have revealed insights into the relative roles played by RF field- and space-charge-induced [61,62] dynamics. Most relevant to our present work, a detailed study of parametric variations was recently completed that aimed at proposing changes to the current LCLS photoinjector [46]. The study used a genetic optimizer, investigating trade-offs between peak current and emittance, with the goal of optimizing parameters such as the drift length between the gun and post-acceleration linac. The maximum field on the photocathode was kept the same as currently used in the 1.6 cell geometry, at $E_0$=120 MV/m. In this case, the parameter search yielded a similar optimized pulse length (~10 ps) as in the original LCLS photoinjector, but with a smaller radial extent at emission (as $Q_b$ is reduced from 1 to 0.2 nC). The results of this study indicated that at this charge, one obtains a 20 A beam – a factor of 5 lower than the original LCLS injector – but with an emittance of $\varepsilon_n = 110$ nm-rad, or a full order of magnitude lower than the original LCLS design.

The most notable change proposed to the optimized working point compared to the present LCLS approach was to move the position of the post-acceleration linac from *z*=1.5 m to *z*=2.2 m. This is due to a reduction of beam charge from the LCLS 1 nC scenario, which is roughly mid-way between the blowout and cigar limits, to a quasi-cigar beam-like case where one has notable *radial* blowout. In Ref. 46 a relative drop in beam density causes the transverse beam plasma frequency to diminish, and a longer drift length is needed before the completion of the transverse plasma oscillation required for compensation. Note that this longer drift, to *z*=2.2 m, is also what was found in the analysis of the 250 MV/m, low charge cigar-beam case discussed in the previous section. We will exploit this insight further when we return to the S-band case in the next section.

We can immediately profit from this proposed LCLS injector optimization by using the well-established scaling methods developed in Ref. 45t o establish a working point near 250 MV/m by changing the operating RF frequency $f_{RF}$. Scaling with respect to $f_{RF}$ means that all frequencies in the problem, including the spatial rate of acceleration and focusing and the beam plasma frequency must also scale proportionally. This implies that, to scale the optimized LCLS proposal to an RF frequency twice that of S-band ($f_{RF}$ =5.712 GHz in C-band), we should choose $E_0 = 240.$ MV/m, nearly identical to the value assumed for our S-band cryogenic gun. In this case we must also scale the focusing fields up by a factor of two, and shrink all beamline dimensions similarly. To preserve the beam-plasma behavior, we must also scale all the beam dimensions $\sigma_i \propto \lambda_{RF}$ and the charge must scale as $Q_b \propto \lambda_{RF}$. As a result of these scaling laws the beam envelope and emittance evolution are preserved, and the emittance also is known to scale exactly as $\varepsilon_n \propto \lambda_{RF}$.

This approach has been explored in simulations found in Ref. [63]. With 100 pC in a scaled C-band 1.6 RF gun having the same interior shape as the standard S-band device [29], and using post-acceleration (with C-band linacs operated at $E_{acc} = 35$ MV/m) that begins at $z$=1.1 m downstream of the photocathode [64], one achieves an emittance of $\varepsilon_n = 55$ nm-rad, again with 20 A peak current. Thus, a factor of four increased brightness is obtained with the C-band scaled option over the re-optimized LCLS case, as predicted by the methods developed in Ref. 45.

This is a promising result, which indicates possible scaling of an optimized photoinjector design to higher RF frequency using quasi-cigar-beam conditions. This is an area of interest by the current group of authors and their collaborators [63]. We next connect this optimization approach to the focus of this current paper, the S-band case. We study the extension of the quasi-cigar-beam design approach used in this example, as well in the low charge case discussed above, to examine optimized ultra-high brightness beam production in the S-band 1.45 cell RF photocathode gun.

### VII. Optimized working point in cryogenic S-band gun

Following the approach given in Ref. 46, we have computationally scanned the parameter space associated with the cryogenic S-band gun discussed above, *i.e.* a 1.45 cell structure with a peak field $E_0 = 250$ MV/m. The optimum obtained from this analysis entails use of a 200 pC, 10 psec long beam with 1 psec rise and fall times, and a transverse Gaussian distribution cut at $1\sigma$, yielding a rms transverse beam size at emission of 82 $\mu$m. In simulations, the beam is launched at near maximum field, $E_0 \sin \phi_0 = 240$ MV/m. In this case the thermal emittance is $\varepsilon_{th}$=43 nm-rad.

We again find that the main change needed to access this new operating point is found in the drift distance after the gun. The optimized distance to the initial linac section is yet longer, at $z$=2.9 m from the photocathode. This significant lengthening, from the present $z$=1.5 m, reflects the doubling of the energy with respect to the LCLS case, which strongly lowers the plasma frequency. Further, the 1.45 cell geometry does not provide strong transverse focusing just after emission as the 1.6 cell geometry studied above in the C-band scenario does. Thus, the beam plasma frequency is diminished further, and one must wait longer for the emittance compensation process to proceed.

After acceleration through two linac sections of with average gradient 17 MeV/m, the emittance compensation approaches completion, as shown in Figure 7. The final emittance is $\varepsilon_n = 51$ nm-rad, with a slice emittance of 45 mm-mrad; nearly the same value of the emittance is obtained by removing 5% of the beam through collimation. We note that the collimated electrons are indeed found in the head and the tail of the bunch, as suggested by the example of Figure 6 (left).

Note that the peak current in this S-band case remains at ~20 A, as in the C-band case mentioned above. This is also the value found in the modified LCLS scenario studied in Ref. 46. However, due to improved emittance compared to the proposed *new* LCLS photoinjector working point, we find that the predicted brightness is increased by a factor of nearly five (or six with 5% collimation) at the same current. This illustrates quite well the advantages of very high field operation.

In an exercise to explore the limits of electron source and FEL performance, and also to compare with the C-band example, we have scaled $Q_b$ to 100 pC while keeping the emission time $\tau$ constant (peak current $I$=10 A). We have re-optimized the beam optics to minimize the emittance, which is further reduced to $\varepsilon_n$=36 nm-rad. This unprecedented level of performance with moderate charge, corresponding to a brightness of $B_e = 1.6 \times 10^{16}$ A/(m rad)$^2$ – over *two orders of magnitude* larger than the original LCLS design brightness – brings new opportunities in FEL physics. In this regard, we next examine the use of this very bright beam in two X-ray FEL examples. In the first case the beam is compressed and injected into the present LCLS injector for lasing at 1.4 Å;

the second scenario explored concerns exploitation of the lowered emittance to permit robust operation at an extreme hard X-ray wavelength, 0.155 Å. We discuss issues associated with beam compression and emittance preservation, and perform start-to-end simulations of XFEL performance based on an approach that uses a final compression through micro-bunching.

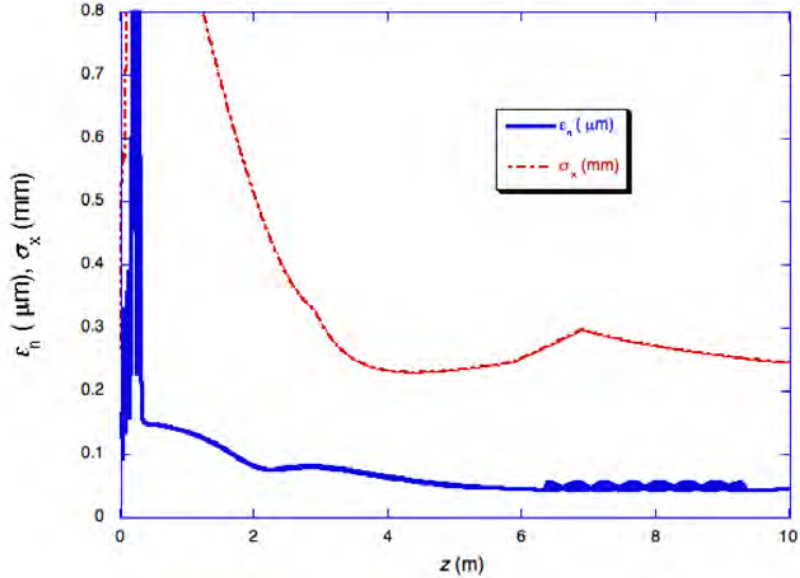

**Figure 7. Evolution of transverse beam size and normalized emittance in S-band photoinjector, with 1.45 cell RF gun operated at 250 MV/m followed by two 3-m traveling wave linac sections, using cigar-like beam with 200 pC charge and 10 ps FWHM bunch length.**

## VIII. Arriving at an X-ray FEL: Physics Issues and Simulation Results

The transverse and longitudinal beam phase spaces obtained at the exit of the photoinjector for the 100 pC case, at a mean energy of 110 MeV, are shown in Figure 8. This beam has excellent phase space qualities, but the current is obviously too low to permit its use in X-ray FELs. At present, the beam in the LCLS is compressed to give currents starting from the few 10's of Amperes after the photoinjector, but reaching the multi-kA level at high (multi-GeV) energy, thus providing strong X-ray FEL gain. The process of transport, acceleration and compression is aimed at enhancement of $B_e$ through an increase in current, while attempting to preserve the emittance.

Just as was confronted in the process of optimizing space charge effects through emittance compensation, collective effects present a challenge in realizing higher brightness through beam compression. There are two major effects limiting $B_e$ between the photoinjector exit and the FEL undulator. The first is revealed when, after reaching moderate energy, the beam negotiates a bend, introducing longitudinal dispersion. This converts longitudinal space-charge-induced energy changes into spatial density modulations, a process that is described as longitudinal space-charge (LSC) instability [65,66]. This effect is manifested by the observation of coherent optical transition radiation (COTR) at beam profile monitors [67]. The coherence of the radiation implies that the beam profile is not imaged in these measurements. Instead, the transverse energy density associated with the beam fields is observed, limiting the utility of OTR diagnostics. To mitigate LSC, one must increase the local energy spread in the beam to the several keV level using a laser heater [68,69,70]. This method becomes urgently needed when dealing with such cold, high phase space density beams such as are produced by the high field injector, *cf.* Figure 8(b).

Micro-bunching and related brightness-reducing phase space distortions arise in bending systems from another collective process, that of coherent synchrotron radiation, or CSR [71,72]. This mechanism is an instability similar to the FEL itself, with the beam self-organizing due to interaction with its own coherent radiation. It is often studied with the code Elegant [73], as in the start-to-end simulations discussed here. One may see from the longitudinal phase space at injector exit, shown in Figure 8(b), that the slice energy spread before laser heating is indeed very small. In order to stabilize the beam against CSR instability, one also employs a laser heater, typically giving an incoherent energy spread of a few keV for the LCLS-like scenarios of interest here.

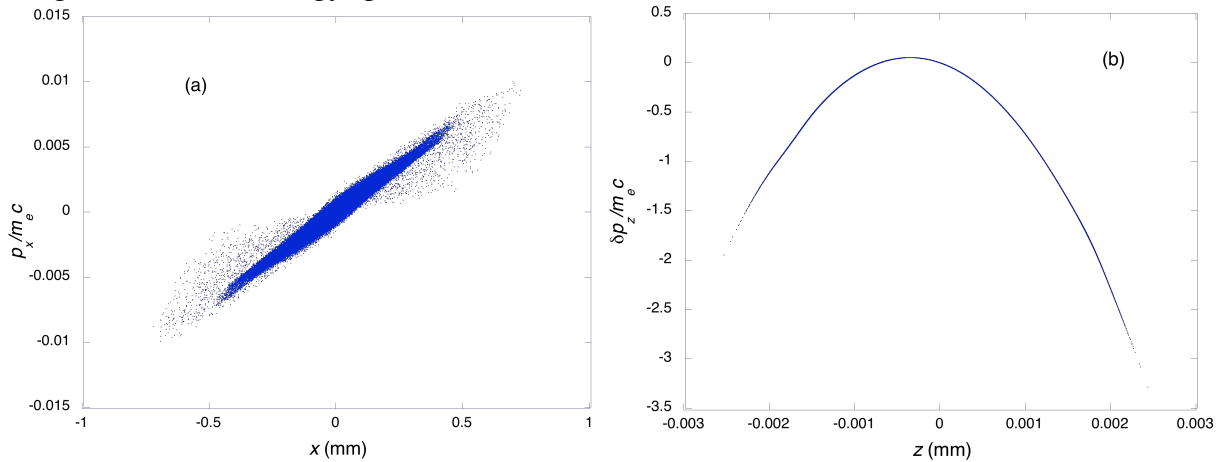

Figure 8. (a) Horizontal phase space for at end of injector, 100 pC start-to-end simulations; (b) longitudinal phase space for same beam before laser heater.

When compressing the longitudinal phase space of the 100 pC beam introduced above, growth in energy spread due to macroscopic CSR, which cannot be suppressed by the laser heater, remains a challenge to exploitation of an ultra-high brightness source. This is both because of the growth in energy spread and concomitant emittance growth due to uncancelled dispersion. In Elegant simulations that use the phase space given in Figure 8 as input, we find the onset of non-negligible growth in $\varepsilon_n$ near 1 kA for the two-compressor transport used, despite amelioration of CSR effects using of a dispersion management scheme introduced in Ref. [74].

There is a further fundamental challenge limiting the effectiveness of compressing a beam for FEL while simultaneously preserving the beam phase space qualities needed for lasing – restricting the final slice energy spread. As in practice the laser heater sets the energy spread of the beam before compression (set, *e.g.*, in the example below to 3 keV), the 6D brightness is proportional to the 5D high brightness we have been discussing, multiplied by the relative energy spread induced. The degree of compression demanded increases the initial slice energy spread by the same factor, thus giving an advantage to producing short beams at the source. In this way if one aims at a given peak current, the final 5D brightness is approximately proportional to the 5D source brightness. We note that the achieving of low emittance through launching longer beams (and thus lowering the source brightness) can also bring about a different problem, the need to correct nonlinear correlations of the longitudinal phase space through use of RF harmonics during acceleration [75], as well as other methods [76,77,78]. With shorter, higher brightness beams, this challenge can also be mitigated.

While obeying the constraint on final energy spread in reaching the desired currents, we must simultaneously avoid CSR-induced emittance growth. We thus study here a scheme that avoids compressing the beam as a whole, but instead employs a concept termed ESASE [40] where beam micro-bunching at the μm scale is induced through an inverse free-electron laser (IFEL)-based bunching section. Therefore, after using an Elegant simulation corresponding to current LCLS beamlines, including acceleration to 14.1 GeV, with the laser heater and two conventional chicanes, we introduce a micro-bunching system based on a 2 μm laser, as used in the current ESASE demonstration experiment at the LCLS, XLEAP [79]. Simulation of the ESASE system's IFEL and chicane was performed using Genesis 1.3 and Elegant. This study produces the nearly 10 kA peak longitudinal current profile (over a wavelength in the center of the beam) shown in Figure 9(a). The rms energy spread in this case is approximately 1.3 MeV, or less than $10^{-4}$ fractional spread. This energy spread is well below that needed to enable FEL gain in the LCLS undulator.

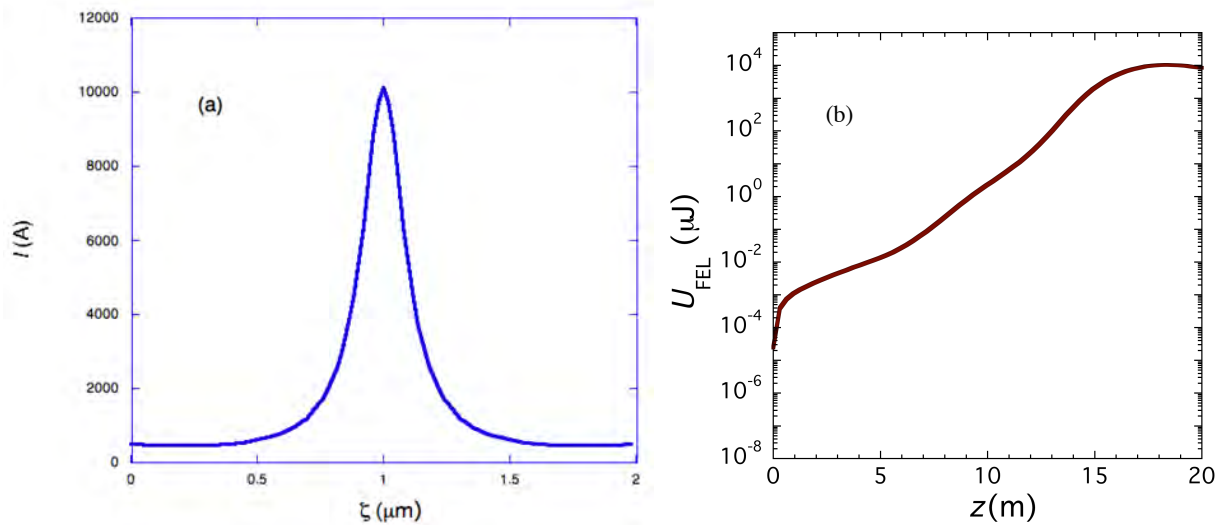

**Figure 9. (a) Current profile after micro-bunching section in ESASE scheme for LCLS parameters. (b) FEL energy for full beam having micro-bunch current as in (a), with central wavelength 1.41 Å (8.8 keV photons).**

This micro-bunched beam's phase space is utilized in Genesis 1.3 to study the X-ray FEL performance. As a benchmark, we first examine the use of this beam in the present LCLS undulator, which has a period $\lambda_u$=3 cm, and undulator parameter $K$=2.475, yielding fundamental wavelength of 1.41 Å. In this case the lasing proceeds quite rapidly, with a gain length ~70 cm and saturation achieved in ~17 m. The FEL cooperation length (3.5 nm) is, we note, much shorter than the FWHM microbunch length of ~200 nm. These values are a factor of three-to-four smaller than in the LCLS, due to use of a brighter beam. The total photon pulse energy is 6.9 mJ, which despite the fact that only a 100 pC charge is used, is also a factor of around three times higher than in nominal LCLS operations. This result is straightforwardly extended to use of 200 pC beams, which have slightly higher emittance, but similar performance, yielding over 11 mJ. We note that this system can be optimized further by focusing the beam to a smaller size – in this case the beta-function is near 10 m, and this can be made smaller to increase the gain. Further, this undulator may be tapered within limits arising from slippage to increase the power extraction efficiency. These topics are beyond the scope of the present work, but are currently under investigation.

Given significantly higher brightness, we see that the gain and associated efficiency are strongly enhanced by use of beams derived from the cryogenic high field photoinjector. With such a small emittance (that is well preserved by the ESASE scheme), we can also explore the short wavelength frontier of X-ray FEL operation. To this end, we take the undulator as described in Refs. [41] and [80], which is a Pr-based cryogenic device having period $\lambda_u$=9 mm, and strength $K$=1.8. We note that this undulator has a narrow gap, which introduces problems with resistive wall impedance [81]; this is mitigated by the use of microbunching in the ESASE approach. This scenario yields excellent FEL coupling at 0.155 Å (80 keV), with a predicted 3D gain length of 77 cm [82] (implying $\rho = 5.3 \times 10^{-4}$), a result borne out in the simulation shown in Figure 9(b). The total energy radiated including all micro-bunches is 540 µJ. This corresponds to $4 \times 10^{10}$, photons at a wavelength one order of magnitude shorter than currently available, dramatically illustrating the capabilities in FEL enabled by this new class of electron source.

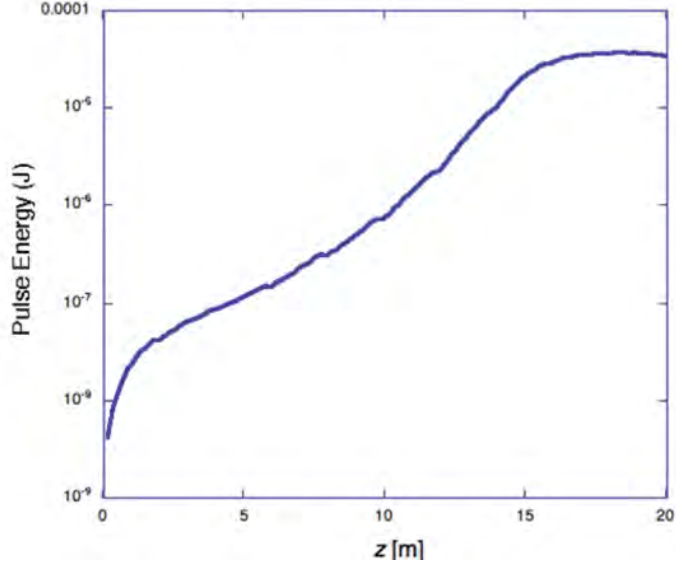

**Figure 10. Simulation of FEL energy evolution with beam current profile and transverse phase space as in Figure 9(a), and $\lambda_u$=9 mm, undulator, and 14.1 GeV beam energy, lasing with central wavelength 0.155 Å (80 keV photons).**

Indeed, the performance predicted in this case would meet or exceed that corresponding to the demands of the MaRIE X-ray FEL project, now entering its preliminary phases at Los Alamos [83,84]. In obtaining short wavelengths, the central advantage is found in the low emittance, which is needed to meet the demands of the criterion $\varepsilon_n \leq \lambda_u \gamma/4\pi$, a limit which is approached at the wavelength evaluated. For MaRIE, which is projected to initially operate at 42 keV, this emittance is more than adequate, and in fact provides a safety margin. The high brightness of this beam provides for the impressive gain needed to produce a compact (*i.e.* short gain length) X-ray FEL. The brightness also yields high efficiency, which may also be enhanced by tapering. MaRIE type scenarios and attendant technical challenges are also explored in Ref. 63.

### IX. Photocathode Performance at Low Temperature and Ultra-cold Beam Emission

As with the surface resistance properties of the gun structure's metallic walls, the emission properties of metallic photocathodes change in advantageous ways at cryogenic temperatures. In metal photocathodes, one can adjust the MTE by tuning the photon energy used to illuminate the cathode to just above its work function [21]. In practice this requires accepting the decrease in quantum efficiency QE in exchange for smaller emittances [85]. Dowell and Schmerge [86] have shown that well-above photoemission threshold, where $h\nu \gg \phi_{eff}$, the photoemission temperature scales as $k_b T_c = (h\nu - \phi_{eff})/3$, and the quantum efficiency obeys $QE = \frac{N_{e-}}{N_\gamma} \propto (h\nu - \phi_{eff})^2$. In this regime, Cu photocathodes typically display $k_b T_c$ ranging from ~100 meV to 1 eV [86, 87], depending on the wavelength used.

Near threshold the situation is very different, as both the occupation of accessible electrons in the metal as well as their associated spread in emission energies are determined by the tail of the electrons' Fermi-Dirac distribution, as analyzed in Refs. [88] and [89]; the model predictions for Cu are plotted in Figure 11. In the limit that $h\nu \to \phi_{eff}$, the photoemission temperature approaches the physical cathode temperature, $k_B T_c \to 26$ meV at 300 °K. This temperature limit has recently been demonstrated for an antimony photocathode in a DC emitter [89]. Cooling the cathode to from room temperature to below 30 °K reduces the possible minimum $T_c$ by an order of magnitude.

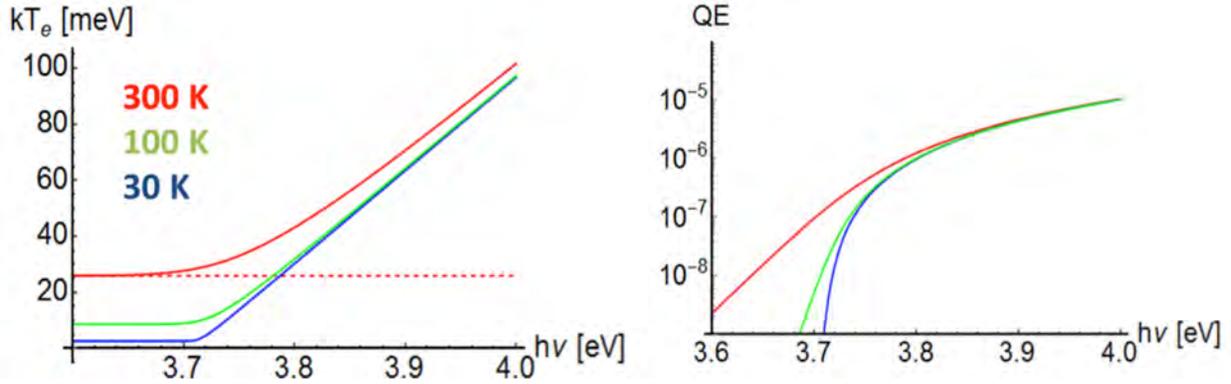

**Figure 11. Photoemission temperature $k_B T_e$ (left) and quantum efficiency QE (right) as a function of photon energy, for atomically clean Cu [55] according to the relations given in [57, 58], using an applied field of 250MV/m. The scale of the quantum efficiency curve is such that at zero field, 270 nm photons produce a quantum efficiency of $10^{-5}$, near what was attained in [90] . The energy at 300 K is shown by the dotted red line.**

The exploitation of low emission temperatures is complicated by the presence of large fields due to the applied laser and RF power, as well as the self-fields of the emitted beam. To illustrate the challenges associated with achieving low $k_B T_c$, we examine Cu photoemission initiated by 3.73 eV photons. This corresponds to a QE of $\sim 5 \times 10^{-8}$, which requires 130 µJ in a 2 psec laser pulse to generate a 1.67 pC electron beam, as described above. The minimum $k_B T_c$ for this wavelength is ~7 meV. The laser flux implied is above the damage threshold of Cu for this pulse length, however. In fact, to operate without damage [91], one should decrease the laser intensity by an order of magnitude, implying a factor of 3 increase in $k_B T_c$ to ~21 meV. The associated change in $h\nu$ also yields an increase in $k_B T_c$, giving $\varepsilon_{th} = 1$ nm rad, still a factor of 5 smaller than that at 300 °K. One may avoid flux limits with a high QE, multi-alkali semiconductor photocathode [92, 93]. Such photocathodes have ~psec time response [94], allowing operation in the cigar-beam limit.

Achieving notably smaller emission temperature in a cryogenic, high field scenario will require overcoming a number of effects. First, we note that the maximum work function lowering due to the Schottky effect [90,95] is large for high fields, $\Delta\phi_{eff}(eV) = 0.038 \sqrt{eE_t(MV/m)}$. Here $E_t$ represents the total longitudinal electric field at emission, including space charge. For our parameters, the maximum $\Delta\phi_{eff}$ is high, ~0.59 eV. This can be compensated by operation at a longer laser wavelength. However, the strength of the Schotty effect varies along the beam during emission; if left uncompensated this would cause a time-varying effective temperature $T_c$. Keeping in mind that operation near $Q_{b,\max}$ is inadvisable due to strong bunch lengthening, one can assume the maximum fractional change in $E_t$ due to space charge-derived fields at the cathode (see

Appendix A) is $\alpha_{sc}$ <0.2. In this scenario the Schottky potential lowering variation is ~$\alpha_{sc}/2$. For our blowout regime example this corresponds to >100 meV change in the Schottky work function lowering. Investigations into the clarification and mitigation of the Schotty effects are ongoing.

The need to consider the proximity of laser-induced damage threshold in metals indicates that a larger temperature $T_c$ may arise from fast local electron heating in the photocathode from high intensity laser illumination. The time scales of this process are regulated by the relaxation of the electronic temperature to the metal lattice, which is material and laser fluence dependent [96]. This effect is important for very short pulse length lasers, and occurs in scenarios where one should also consider multi-photon emission [97,98]. Both phenomena and their effects on the MTE have recently been studied experimentally [99] and theoretically [100]. While for very short (<50 fsec) laser pulses, these effects are serious, they are less so for the relatively relaxed cigar-beams. Even so, unless one uses a semi-conductor photocathode, the inherent MTE from multi-photon and laser-induced heating may limit the MTE to a few 100 meV [99]. Regarding RF-derived heating, in the highest dissipation region, the surface temperature rise is below 10 °K. However, photoemission is limited to a region close (<200 μm) to the axis where the RF dissipation is negligible and there is ignorable impact on the emission characteristics.

At such small temperatures, the laser bandwidth also may play a significant and potentially useful role. For an 8 fs transform limited Ti:Sapphire laser pulse, the spectrum contains ~300 nm FWHM bandwidth, or a 580 meV spread in energy. In the 1.6 pC cigar case we have considered above, the photocathode drive laser pulse length is 2 ps full width, implying that if we utilized the 8 fs transform limited pulse in chirped-stretched mode, there is a linear correlation between photon energy and emission time. This correlation may be used to compensate the linear component of the time dependence of the Schottky-induced potential component due to space-charge. One must also consider the effects of photocathode surface imperfections on $T_c$. Surface cleanliness has been shown to change the QE of metallic photocathodes [86] by more than an order of magnitude, which in turn demands adjustment of the laser fluence. Beyond this, lack of uniformity in both the work function and surface roughness can have a significant impact on beams with small $k_B T_c$. Both are areas of active research in photocathode physics [101][102][103].

### X. Radiofrequency Cavity Surface Resistivity, Quality-factor and Coupling

With the improved beam emission and dynamics performance and their implications for applications discussed, we now analyze critical aspects of using cryogenic, high-field Cu cavities. As noted above, the advantages conferred by cryogenic cavities in high field operation arise from enhanced material hardness, smaller coefficient of thermal expansion, and lower surface dissipation, with concomitant mitigation of pulsed heating stresses. The improved yield strength of the metal enables very high fields to be reached, while the mode of dissipation and structure expansion dictates important design features of the RF cavity system.

To appreciate the experimental investigations presented below, we first give some theoretical background. The lowering of the surface power dissipation at low material temperature $T_m$ was initially investigated by London [104], who found that surface resistances $R_s$ in metals at MHz frequencies and low temperature are not accurately predicted by the classical model based on the conductivity following Ohm's law. The theory explaining this phenomenon, which is termed the anomalous skin effect (ASE) of metals, was then developed by Reuter and Sondheimer [105]. As this theoretical work is well established, we recapitulate only the relevant results here.

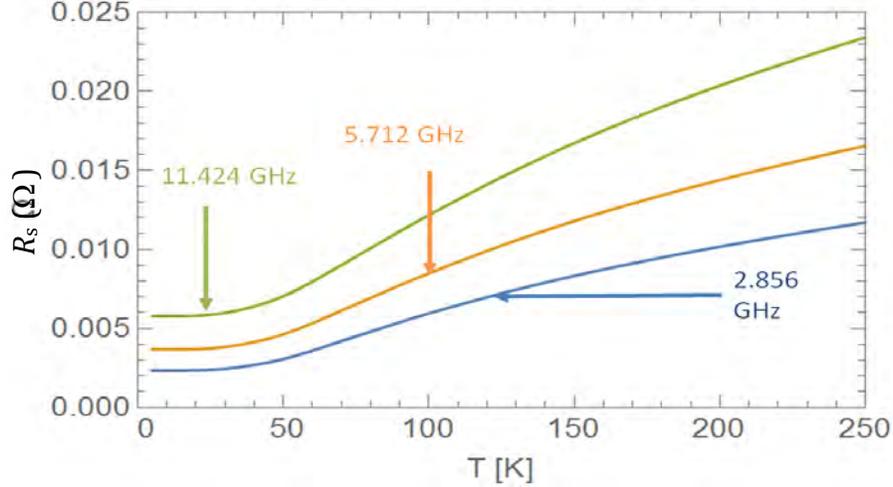

Figure 12. Comparison of the anomalous skin effect surface resistance in RRR=400 Cu at 2.856, 5.712, and 11.424 GHz.

In the case of ohmic dissipation, the surface resistance is found from the complex impedance $Z_s(\omega) = \sqrt{\frac{\omega Z_0}{2c\sigma_c}}(1-i)$, where the $Z_0$ is the impedance of free space, as $R_s(\omega) = Re[Z_s]$. Thus $R_s$ should depend on the ratio square root of the ratio of $\omega = 2\pi f_{RF}$ to the ohmic conductivity $\sigma_c$. At cryogenic temperatures, $\sigma_c$ is two-to-three orders of magnitude larger than at ambient temperature. At very low $T_m$ and high $\omega$, deviations from ohmic behavior are expected. As the metal's temperature decreases the mean free path of electrons increases as indicated by $l_f = \frac{\sigma_c Z_0 c v_f}{\omega_p^2}$, where $\omega_p$ is the plasma frequency, and the collision frequency $v_f$ increases as $T_m^{-1/2}$. This behavior is evidenced by the decreasing DC bulk conductivity [106] at lower temperature. As $T_m$ decreases, $l_f$ and the electromagnetic skin depth $\delta$ become equivalent in scale. Ohm's law requires that the electric field in the conductor does not vary over the free path of the electrons, but at low enough temperature the electric field changes within a spatial scale of $\delta$, and this assumption no longer holds. To find the current density, the electric field must to be integrated over the path of the electrons in the metal as the response of conduction electrons varies on the scale of the mean free path, as has been done in Ref. 105. Thus, one may not express the current density simply in terms of a bulk ohmic conductivity [107]. A careful analysis of the ASE yields a notably different surface resistivity dependence on material and wave properties than ohmic behavior predicts. While the expression for the impedance is not easily reduced, in the low $T_m$ limit one may write

$$Z_s(\omega) = Z_0 \left[\frac{\sqrt{3}v_f}{16\pi c}\left(\frac{\omega}{\omega_p}\right)^2\right]^{\frac{1}{3}}(1-\sqrt{3}i). \tag{9}$$

It can be seen that the surface resistance depends on the frequency as $\omega^{2/3}$. The surface resistance of Cu as a function of measured $T_m$ (indicated simply as $T$) is given in Figure 12 for three different RF frequencies: $f_{RF}$=2.856, 5.712, and 11.424 GHz assuming a RRR=400 (residual-resistance ratio, the ratio of bulk resistivity at 300 °K to that in the $T=0$ limit). The switch from $\omega^{1/2}$ to $\omega^{2/3}$ scaling is apparent; instead of a factor of 2 difference between S- and X-band cases, the ratio found in $R_s$ is ~$4^{2/3}$=2.51.

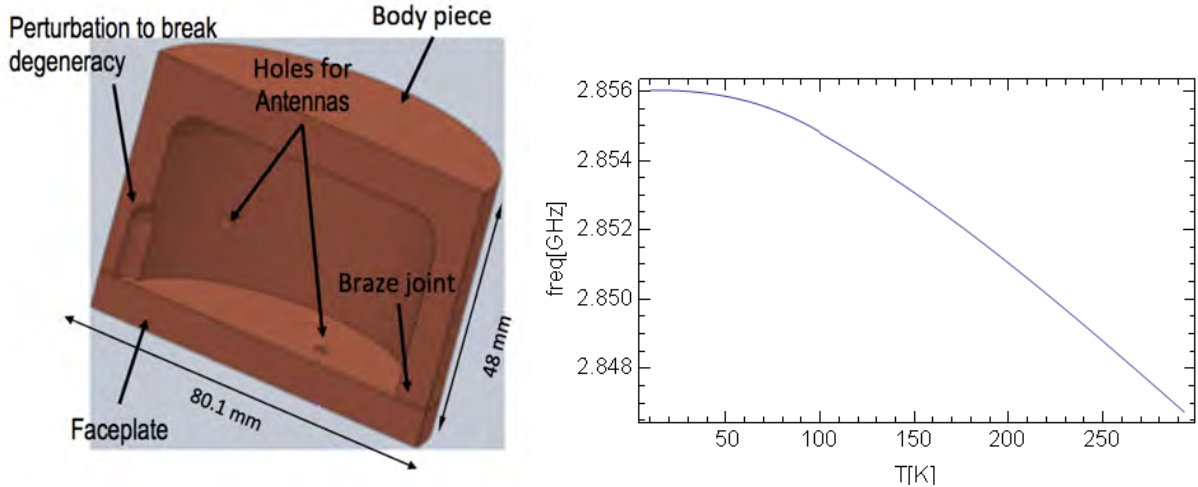

Figure 13. (left) Cutaway model for section of S-band Cu pillbox test cavity. The faceplate is brazed to the bottom, while on the left a feature is included to break dipole mode degeneracies. Through-holes host two antennas for $S_{11}$ and $S_{21}$ tests. (right) Calculation of temperature dependence in the $TM_{010}$ accelerating mode using the thermal coefficient of expansion for Cu.

To provide essential experimental input for the design of the S-band RF gun and its coupling system, an investigation of the cryogenic properties of an oxygen-free high conductivity Cu test cell was performed. The test cavity has a pillbox form that is manufactured from two pieces of copper, with the body of the cavity and a faceplate that is brazed on, as shown in Figure 13 (left).

The geometry of the pillbox cavity shown was chosen so that the $TM_{010}$ mode resonant frequency is equal to 2.856 GHz at 20 °K. Figure 13 (right) also shows the prediction of the $T_m$-dependence of the mode resonant frequency for the given Cu material properties. At room temperature the cavity $TM_{010}$ mode is found at 2.847 GHz. The ambient temperature value of $Q_0$ $Q_o$ in the test cavity was calculated to be $1.8 \times 10^4$, while the cavity external coupling factor was tuned via the length of the antenna to obtain $\beta_c = 0.3$, so that when $R_s$ decreases by a factor of ~5 at cryogenic temperatures, the coupling is close to critical.

Two versions of the cavities were manufactured from different Cu stock and at two different machine shops, located at UCLA and SLAC, respectively. Data was taken on both cavities at a range of temperatures from 300 °K down to 4 °K using a cryo-mechanical refrigerator-cooled cryostat at SLAC. The internal quality factor $Q_0$ was measured every 0.1 °K as the cryostat warmed to room temperature. In Figure 14, the data from these scans compared to the theoretical value of $R_s$ in Cu with RRR=400 and IACS of 95% is shown. Here both RRR and IACS are taken from the relevant material data sheets. The measured $R_s$ displays the expected behavior, but with a slightly degraded value of the warm-to-cold ratio of $Q_0$; it is found to be 4.63 as opposed to the expected 5.4. This is likely due to an incomplete knowledge of the material RRR.

The external coupling of the RF photoinjector cavities must be chosen to balance the competing priorities of achieving 250 MV/m peak field, and minimizing the total power dissipated in the structure. We assume for the calculation of operating parameters that the structure will be used at 27 °K, with liquid Ne used as coolant. Liquid neon has a heat capacity 40 times that of liquid helium, and thus is very useful for cooling despite the narrow 3-degree range in which it occupies the liquid state. Further, 27 °K is an ideal operating temperature for cryogenic Cu, as it is below the knee in the $R_s(T)$ curve, small heating effects during the RF pulse do not notably change the surface dissipation properties. Further, the coefficient of thermal expansion is very small at these temperatures. The resistance to thermal changes in the RF structure response is thus quite robust.

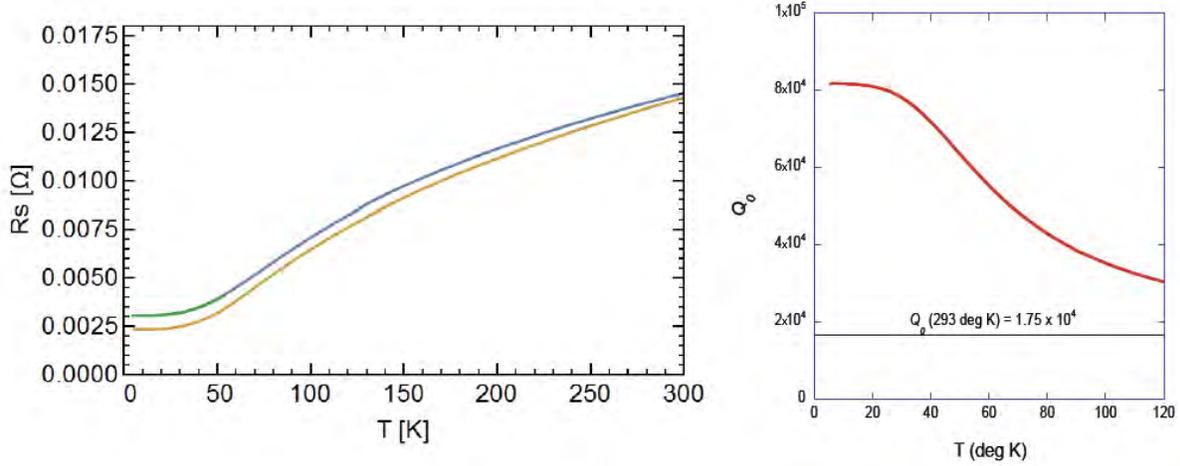

**Figure 14.** (left) RF surface resistance of both accelerating cavities, SLAC (blue) and UCLA (green). This is compared to a theoretical RF surface resistance of copper with IACS 95% and RRR=400. (right) Quality factor in SLAC test cavity.

We are now in a position to outline the parameters of the external coupling scheme. The 1.45 cell gun has 7.2 cm of active length, and is fed by 50 MW, below the standard output of a SLAC S-band 5045 klystron. The structure is highly over-coupled at cryogenic temperatures in order to input and remove RF power quickly. In addition, phase reversal of the drive is used to empty the RF gun cavity in a short time, further minimizing the total RF power dissipated. The parameters of the RF coupling and gun system are given in Table 2.

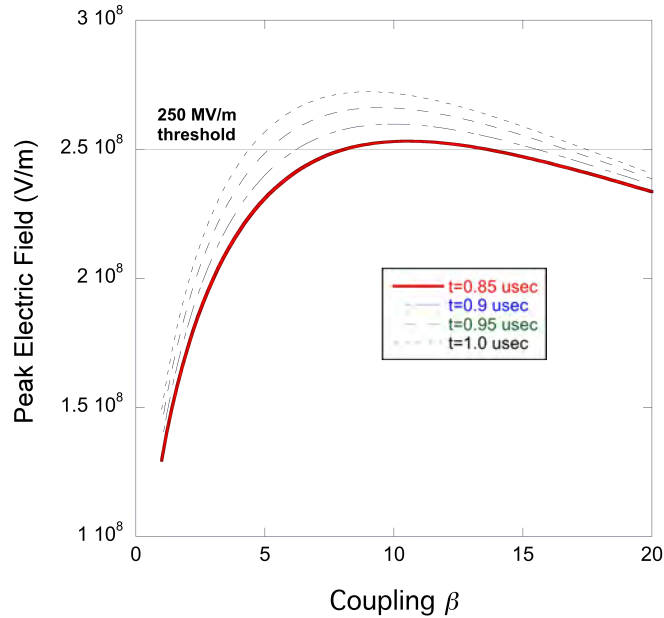

**Figure 15.** Peak photocathode electric field as a function of $\beta$ in 1.45 cell RF gun, for 4 different RF pulse lengths.

Under these assumptions, we examine the conditions under which one may reach $E_0$=250 MV/m at the photocathode. In Figure 15 we show the dependence of $E_0$ on the coupling used, employing four different RF pulse lengths: $\tau_{RF}$=0.85, 0.9, 0.95 and 1.0 μsec. It can be seen that the goal of 250 MV/m is comfortably reached for $\tau_{RF}$=0.9 μsec using a coupling $\beta$=9. Taking this as the design coupling parameter, with $\tau_{RF}$=0.9 μsec the total energy dissipated per RF pulse is determined to be 3.04 J, giving a power load at cryogenic temperature of 375 W for 120 Hz operation. We note that the value $\beta_c=9$ at 27 °K implies $\beta=1.95$ at 300 °K, and thus the structure is similar in its coupling geometry to present devices [108].

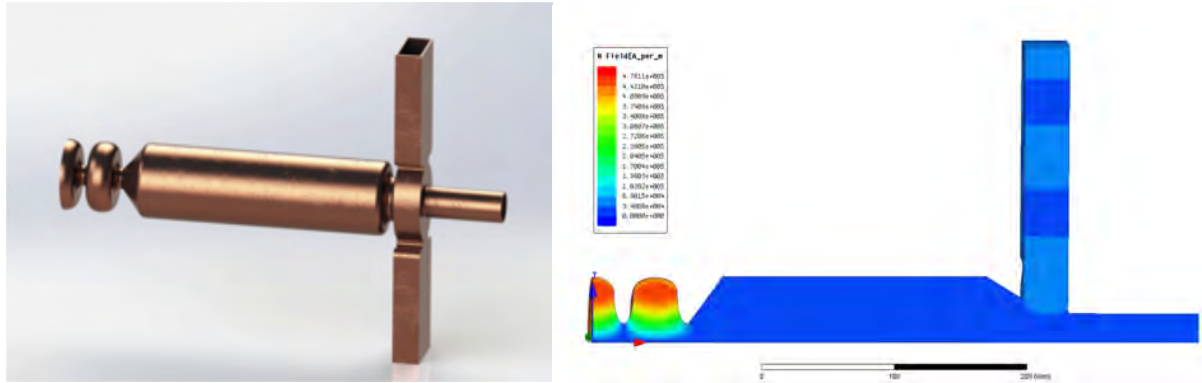

Figure 16. (left) Three-dimensional outside view rendering of RF photoinjector and external power coupling, showing on right symmetrized waveguide feeding a mode-launcher style coupler. Cylindrical waveguide then axially couples power into 1.45 cell gun structure. (right) surface magnetic fields in coupler and gun, as simulated in HFSS, with color map extending from 0 (blue) to 5x10$^5$ A/m (red).

We illustrate the basic layout of the RF gun and coupler system in Figure 16. Here we show the RF photoinjector and external power coupling system, which employs a symmetrized waveguide feeding a mode-launcher style coupler [109] which transports the power towards the gun through a cylindrical waveguide. This power is axially coupled into the 1.45 cell gun, exciting the desired π-mode, as shown in the HFSS simulation [110] in Figure 16. This power must be removed using a cryo-cooler, as indicated schematically in Figure 1. Investigations of the availability of such a cooler operating at 27 °K have revealed that the efficiency of such a cooling device, beyond the Carnot factor of 0.09, would be ~0.12. This implies that the total wall plug power of the cryo-cooler should be nearly 35 kW. This is a challenging but feasible level of cryogenic power to handle with existing technology, for example Stirling cycle cryo-generators.

Table 2. Parameters of RF gun and feed system for study in Figure 19. The last two entries assume $\beta=9$ and $\tau=0.9$ μsec.

| Internal quality factor $Q_0$ (300 °K) | 13,483 |
|---|---|
| Internal quality factor $Q_0$ (27 °K) | 62,425 |
| Input power | 50 MW |
| Normalized shunt impedance $R/Q$ | 136 Ω |
| Peak field at end of RF fill | 250 MV/m |
| Fill time ($\beta_c=9$) | 0.9 μsec |
| Energy dissipated/pulse ($\tau=0.9$ μs) | 3.04 (365 W at 120 Hz) |

## XI. Prospects for higher RF frequency, higher field photoinjectors

We have seen, through the above analysis, the advantages to RF photoinjector performance, in terms of a dramatically lowered $\varepsilon_n$ and associated increase in $B_e$, provided by operation at the extremely high electric fields now in reach through cryogenic operation. In attempts at reaching similarly large fields in previous photoinjectors, the RF frequency has most often been chosen to be high, as this permits fast RF pulses and minimization of pulsed heating. Indeed, as we have seen in S-band, even using a highly over-coupled system and 50 MW of input power, the minimum optimized RF pulse length is ~0.9 μsec, which leads to high levels of heat dissipation. On the other hand, the nominal cryogenic scaling of fill time as $\tau_{RF} \sim f_{RF}^{-5/3}$ permits, *e.g.*, C-band systems to operate with pulses near 300 nsec. Further, considering a constant $E_0$, the power needed to drive a structure of scaled geometry is smaller by $P \sim f_{RF}^{-2}$. Thus, higher RF frequency mitigates power considerations such as total power usage and associated cooling load. In the S-band case studied above, the long pulse required in a 120 Hz photoinjector dissipates ~365 W at cryo-temperatures, driving ambitious cryo-cooler requirements.

A higher RF frequency system would be much less demanding in this regard, and one may even consider fields higher than 250 MV/m. Faster fill times would also give flexibility in the pulse format, as desired by the MaRIE FEL, for example. Also, at high field values, integrated dark current becomes a potentially significant problem that is exacerbated by long RF pulses. As an example, L-band RF guns can be conditioned to sustain fields of 90 MV/m on the cathode [111] but high field emission with large charge per RF pulse remains. Further, recent tests by the current authors shows that dark current in X-band cavities is significant enough to lower the value of $Q_0$ in the structure through absorption of RF power by the current [112] when $E_0$ >300 MV/m.

This raises a frequency independent issue: operation of an RF photoinjector at ≥250 MV/m demands new approaches to surface treatment or use of coatings such as silicon oxynitride [113] and graphene. Graphene is a promising material, in that it is transparent to light and electron transport [114]. Studies of the deposition of these coatings on Cu substrates will be undertaken in the near future. We are now proceeding to dark current tests based on a needle-enhanced peak field in an RF gun that may access fields up to 700 MV/m [115]. It should be emphasized that dark current suppression is a fundamental issue, entailing resolution of the nature of the field enhancement factor dating back to Fowler and Nordheim [116]. Yet another approach is to remove dark current after the gun by use of strip line kickers that may leave only a few nsec open to beam propagation [117]. This may be necessary, in particular for very low charge operation [36].

If one operates at larger RF frequency, the associated faster RF fill times may also be exploited to operate at higher fields. While this option ameliorates space-charge effects on emittance, there are also contributions to $\varepsilon_n$ arising directly from RF forces as well as attendant energy spread in the beam that scales as $\varepsilon_n \propto f_{RF}^2$, a problem particularly noted higher charge. Wakefields in both single and multi-bunch operation are stronger for larger values of $f_{RF}$.

Nevertheless, a promising beam dynamics optimization is found in C-band [63], where the value of $\alpha_{RF}$ used would be ~2 for at 250 MV/m, and the approach to emittance compensation is familiar, scalable from present S-band designs now commonly employed. Use of C-band may also permit operation up to $E_0$=300 MV/m. Further, given the easing of present limitations on brightness, an initiative has been launched by the current authors collaboration, to apply this method of high field photoinjection to enable asymmetric emittance sources for linear colliders [118] and laser-driven accelerators [119,120].

As noted, there are practical issues in scaling RF photoinjectors to frequencies beyond S-band [63]. The devices are more compact and demand focusing over shorter distances, making realization of solenoids challenging. Higher current densities may, in this regard, be possible with cryogenic operation. Small dimensions also cause difficulties in laser injection, and exacerbate spatial and temporal jitter tolerances. Also, at high power, circulators used to protect the RF power sources are difficult. One may avoid RF reflections by use of innovative gun designs, such as the hybrid standing-travelling wave [121,122] and traveling wave [123] photoinjectors. These solutions are attractive for other reasons, including experimentally demonstrated inherent velocity bunching for ultra-short beam creation at low energy in the hybrid [124]), and short $\tau_f$ in the traveling wave device. It is promising to consider cryogenic operation to permit development of an X-band photoinjector based on the hybrid design, given that previous analyses have assumed that one should use a peak cathode field in the X-band hybrid of ~240 MV/m [121].

### XII. Conclusions

We have presented a detailed analysis of the use of a cryogenic copper structure operated at unprecedented high electric fields for creating an RF photoinjector. In the process, we have investigated the issue of beam dynamics optimization in a variety of regimes in which both the extraction fields and the beam energy at the photoinjector exit are much higher than currently encountered. These dynamics studies survey, through scaling laws and simulations, both blowout and cigar-beam regime operation. In the 1D space-charge limited flow blowout regime, we have produced examples in which very high currents can be produced, as may be needed for wakefield applications. We have also observed that this regime may not produce fully optimized space-charge emittance compensation. We have thus placed emphasis in low and moderate charge beam applications on exploiting advantages of the cigar- and quasi-cigar-beam regimes. We have shown that by use of the cigar-beam and quasi-cigar-beam regimes at high fields one may obtain well over an order of magnitude increase in beam brightness. At very low charge, this performance may give significant improvements in applications such as UED and UEM – where the scenario discussed gives a factor of 50 in increased brightness on recent proposal, permitting much higher temporal resolution in UEM. At intermediate charge, the reduction in emittance strongly and positively affects the outlook for future X-ray free-electron lasers. One may operate this new type of XFEL injector to obtain emittances, with charges at the few 100's of pC level, that are lower than current sources by an order of magnitude.

Just as it was necessary to address the post-emission management of space-charge-induced emittance growth through a revisiting of the emittance compensation process, we must evaluate the methods needed to strongly compress – enhance the current – beams for FEL application that preserve the transverse emittance. After reviewing the physical challenges encountered during chicane-based compression, we have discussed a scenario that employs micro-bunching instead of full-beam bunching. With this strategy we have found a scenario in which the gain and efficiency of an LCLS-like X-ray FEL are both greatly enhanced. We also examined the possibility of reaching an order of magnitude higher X-ray FEL photon energy, with extremely strong gain. This approach may thus enable new X-ray FEL capabilities that can be employed in the context of future projects such as the MaRIE FEL.

The harder X-ray case discussed in this paper proposes use of very short period undulators. One may take this concept a step further and use yet shorter period undulators, to the sub-mm-period scale [125], that may be use for creating X-ray FELs at very low beam energy [126]. This type of free-electron laser demands that very strong focusing [127] and extremely high brightness, low emittance beams be used. Further, the use of a micro-bunched pulse train as discussed above presents notable advantages in managing the problem of resistive wall wakefields in such narrow aperture devices [125]. This approach to an ultra-compact X-ray FEL is now under serious study.

In support of the opening of new applications in FEL and direct electron-based imaging, an example study of the physics and technological aspects of a cryogenic RF photoinjector system has been presented. For this purpose, we have chosen an S-band system that can be straight-forwardly deployed in the existing machines such as the LCLS hard X-ray FEL injector, its near-term upgrades, and in the many other FELs based on similar RF technology. We have in this context explored the underlying physics issues such as the anomalous skin depth effect, including an experimental investigation of cryogenic copper's performance at low power in S-band. We have examined implementation issues such as RF design and related cryo-cooling technology. Prospects for extension of cryogenic high field methods to higher RF frequency and yet higher fields have been reviewed and promising directions identified. In addition, we have discussed the more general subjects of cathode and near-cathode physics issues.

The current work introduces many remaining, interesting experimental topics to investigate. As such, the development of the S-band incarnation of this next generation electron source is currently proceeding, with work concentrated on high power 1.45 cell structure testing to explore field limits, as well as studies of dark current and its mitigation. This effort is proceeding in parallel with very high peak field (500 MV/m) work in X-band, where pulsed heating and dark current-derived beam loading [31,112] are strong effects. This work is intended to prepare the path for a full cryogenic RF photoinjector prototype that verifies the production of extremely high brightness beams. This experimental environment will permit the complex interplay between interdependent factors such as cavity performance, high field photoemission, low intrinsic beam temperature, and dark current management to be addressed. Further explorations of application of cryogenic field enhancement in higher frequency and higher gradient systems, are also under way. Consistent with the historic importance of the RF photoinjector, these developments may impact a wide variety of fields, including ultra-fast relativistic electron microscopes, advanced accelerators, very short wavelength FELs and high energy electron-positron colliders.

**Acknowledgments**

This work supported by the US DOE Office of High Energy Physics through contract DE-SC0009914, the US DOE Office of Science SCGSR Graduate Student Research Fellowship program, DOE/SU Contract DE-AC02-76-SF00515, and the US NSF Award PHY-1549132, the Center for Bright Beams.

**Appendix**

We begin the analysis of 1D limits on transient current generation by assuming illumination of a photocathode with a laser having a time profile given by the normalized function $g(t_0)$, with transversely uniform emission inside a radius $R$. Assuming prompt emission, the photocurrent is

$$I(t_0) = g(t_0) \qquad (10)$$

where $Q_b$ is the total beam charge, and the emission time is characterized by $g_{\max} \sim \tau^{-1}$. We assume that $c\tau \ll R$, so that the beam's electric field is predominantly longitudinal. We note for completeness that the 1D analysis of space-charge effects is valid when $\gamma c \tau \ll R$. However, there is significant longitudinal rearrangement of electrons only when $\gamma$ is close to unity, so in practice one may perform a 1D analysis with confidence considering cases with $c\tau \ll R$.

Including the effects of the cathode image charge the longitudinal force on an electron is found,

$$F_z(t_0) = -eE_0 + \frac{eE_0}{\epsilon_0}\int_{-\infty}^{t_0} \tilde{g}(\tilde{t_0})\,d\tilde{t_0} = -eE_0 + \frac{eE_0}{\epsilon_0}G(t_0) = -eE_0[1 - \alpha_{sc}(t_0)] \qquad (11)$$

Here, $E_0$ is the emission field, and we have defined the function $G(t_0) = \int_{-\infty}^{t_0} \tilde{g}(\tilde{t}_0) d\tilde{t}_0$ as the integrated fractional beam charge emitted ahead of $t = t_0$. We have implicitly assumed that $G$ is only a function of $t_0$, and can therefore be calculated once and for all at emission. This assumption, that electrons do not overtake each other, is termed *laminar flow*, and will be justified later. The quantity $\sigma_b$ is the beam surface charge density. The maximum field associated with a surface charge is $\frac{\sigma_b}{\epsilon_0}$, and so we normalize the value of the space-charge field through $\alpha_{SC} = \frac{\sigma_b}{\epsilon_0 E_0}$, . In practice, one operates with $\alpha_{SC} \ll 1$; practically, with $\alpha_{SC} < 0.2$ we may obtain nearly uniform current density in the blowout regime. In this analysis, however, we leave the treatment open to both perturbative ($\alpha_{SC} \ll 1$) and non-perturbative ($\alpha_{SC} \leq 1$) cases. In the limit $\alpha_{SC} \simeq 1$, the image charges dominate the physics, and their effects cause strong diminishing of the current obtained at the beam's tail and, eventually, suppression of electron emission from the photocathode.

Under these assumptions we can write the normalized energy of a given electron emitted at $t_0$ as

$$\gamma(z, t_0) = 1 + \gamma'(t_0) z \quad , \tag{12}$$

where

$$\gamma'(t_0) = \frac{F_z(t_0)}{m_e c^2} = \gamma'(t_0)(1 - \alpha_{SC}) G(t_0) \text{ and } \gamma'(t_0) = \frac{|eE_0|}{m_e c^2}. \tag{13}$$

Once the energy is known, one may find the velocity, and integrate it to find $z$ as a function of $t$,

$$c[t(t_0) - t_0] = \int_0^z \frac{d\tilde{z}}{\beta(\tilde{z}, t_0)} = \frac{1}{\gamma'(t_0)} \int_0^{\gamma(z,t_0)} \frac{\tilde{\gamma} d\tilde{\gamma}}{\sqrt{\tilde{\gamma}^2 - 1}} = \frac{1}{\gamma'(t_0)} \sqrt{[\gamma'(t_0) z]^2 - 2\gamma'(t_0) z}. \tag{14}$$

After the electron is relativistic, the relative longitudinal motion slows to give an asymptotic form of the final time,

$$ct_f(t_0) = z + ct_0 + \frac{1}{\gamma'(t_0)} - \frac{1}{\gamma'_0}. \tag{15}$$

or dropping the dependence on position of the measuring point $z$,

$$ct_f(t_0) \simeq ct_0 + \frac{\alpha_{SC} G(t_0)}{\gamma'_0 [1 - \alpha_{SC} G(t_0)]}. \tag{16}$$

Equation 16 may be used to deduce the form of the final beam distribution. Conservation of probability yields that $J$ expands by the factor $\partial t_0 / \partial t_f$, and the final current density is given by

$$J(z, t_f(t_0)) = \frac{g(t_0) \sigma_b}{\partial t_f / \partial t_0}, \tag{17}$$

where, under our assumptions, we may write the differential time mapping as

$$\frac{\partial t_f}{\partial t_0} = 1 + \frac{\alpha_{SC} G(t_0) g(t_0)}{\gamma'_0 [1 - \alpha_{SC} G(t_0)]^2}. \tag{18}$$

Note that in Eqs. 17 and 18 we implicitly are inverting the relationship between the initial and final time coordinates, *i.e.* when we write $t_0$ we imply $t_0(t_f)$. We will not need to write out this relationship until later. Note also that "wave-breaking" or loss of laminarity is given by the condition $\frac{\partial t_f}{\partial t_0} = 0$, which is not allowed inside of the beam ($g > 0$); the assumption of laminarity is thus validated. The current density deduced from Eqs. 17 and 18 is

$$J(z, t_f(t_0)) = \frac{g(t_0) \sigma_b}{1 + \frac{\alpha_{SC} G(t_0) g(t_0)}{\gamma'_0 [1 - \alpha_{SC} G(t_0)]^2}}, \tag{19}$$

which, assuming significant expansion ($\alpha_{sc} \gg c\tau\gamma_0'$) and charge well below maximum ($\alpha_{sc} \ll 1$, approaches a constant value given by

$$|J| \simeq \frac{e\epsilon_0 E_0}{m_e c}. \qquad (20)$$

This is the maximum current obtainable in the 1D limit, as is used in Eq. 2; inspection of Eq. 17 indicates that it is a monotonically decreasing function of $\alpha_{sc}$. It is useful to recast this result in terms of the total current for this uniform density case, with emission up to a hard-edge radius $R$

$$I \simeq \frac{I_0}{4}\left(\frac{eE_0 R}{m_e c^2}\right)^2. \qquad (21)$$

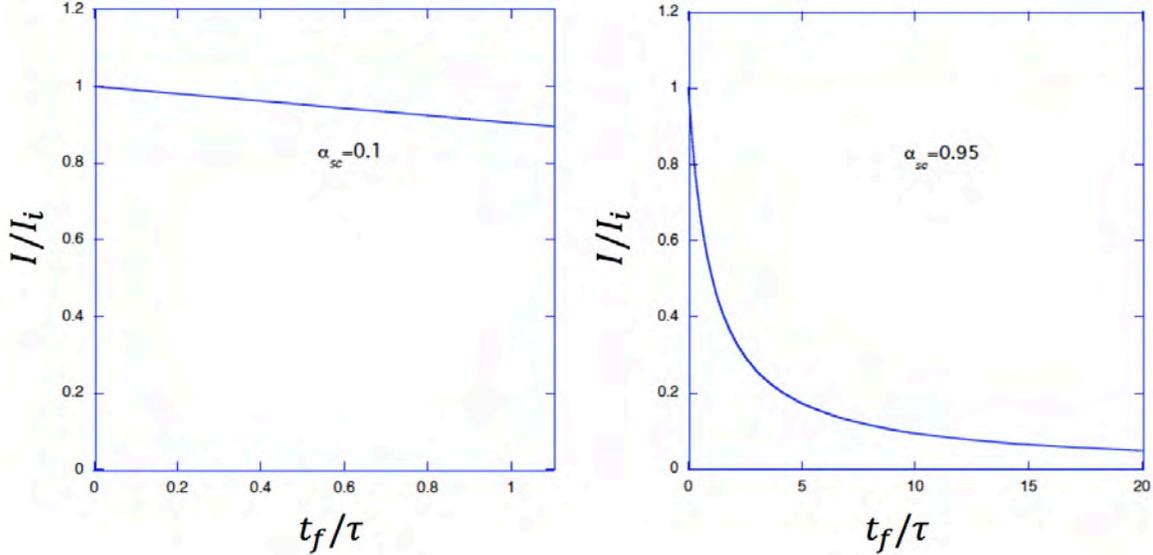

Figure 17. (left) Current (normalized to uniform launch current $I_i$) as a function of final time $t_f$ within the pulse of initial length $\tau$, in the linear limit of emission, $\alpha_{sc} = 0.1$; (right) current profile strong bunch lengthening limit, showing near suppression of emission with $\alpha_{sc} = 0.95$.

A variant of the linear result given in Eq. 20 has been available in the literature for some time [22]; it is indeed the physics basis of the formation of the uniformly filled ellipsoid in the longitudinal blowout regime. It is, as can be seen, obtained from the exact asymptotic analysis of 1D motion under space charge. In Ref. 60 a different scaling for the 1D current limit as a function of the injection field is presented for the longitudinal blowout (termed "pancake" limit therein) regime. That result, is obtained taking the maximum possible image charge forces (cut-off) and assuming that the process of pulse length $\tau$ expansion is arrested by the onset of two-dimensional effects that assert themselves when $c\tau > R$. This would certainly be true for $Q$ approaching $Q_{b,\max}$, but is not so for beams where $\alpha_{sc} \ll 1$, where Eq. 19 applies. It would be necessary to self-consistently merge the present analysis with that of Ref. 60 to give the general limiting behavior of the current in blowout regime as $\alpha_{sc}$ approaches unity.

Here we have presented an extension to the 1D analysis, in which 1D behavior is maintained in the non-perturbative limit. This nonlinear result is obtained by relaxing the assumption $\alpha_{sc} \ll 1$, and it serves to show the effect of diminishing current as more charge is emitted. Indeed, even in the perturbative $\alpha_{sc}=0.05$ case shown in Figure 10, there is a notable sag in the current towards the back of the pulse. In the non-perturbative case where $\alpha_{sc}$ approaches unity (0.95), also shown in Figure A.1, a dramatic pulse lengthening (a factor 20) occurs, accompanied by a strong, non-linear diminishing of the current along the length of the pulse. This type of expansion inevitably would cause 2D considerations to be needed in the analysis. The blowout regime current limit proposed in Ref. 60 gives a mechanism for possible saturation of this longitudinal expansion process.